\documentclass[a4paper,11pt]{article}

\usepackage{amsmath}
\usepackage{afterpage}
\usepackage{booktabs}
\usepackage{bm}
\usepackage{bbm}
\usepackage[justification=justified]{caption}
\usepackage{float}
\usepackage{fullpage}
\usepackage{geometry}
\usepackage[citecolor=blue, urlcolor=blue, linkcolor=black, colorlinks=true]{hyperref}
\usepackage[round]{natbib}
\usepackage{pdflscape}
\usepackage{setspace}
\usepackage{subcaption}
\usepackage{tabularx}
\usepackage[textsize=footnotesize]{todonotes}
\usepackage{xcolor}
\usepackage{comment}
\usepackage[multiple]{footmisc}
\usepackage{verbatim}

\usepackage{fancyhdr} 
\newcommand{\beginappendixA}{%
        \setcounter{table}{0}
        \renewcommand{\thetable}{A\arabic{table}}%
        \setcounter{figure}{0}
        \renewcommand{\thefigure}{A\arabic{figure}}%
     }
\newcommand{\beginappendixB}{%
        \setcounter{table}{0}
        \renewcommand{\thetable}{B\arabic{table}}%
        \setcounter{figure}{0}
        \renewcommand{\thefigure}{B\arabic{figure}}%
     }

\usepackage{graphicx}
\graphicspath{{../graphs/}}

\title{\LARGE{Racial Sentencing Disparities and Differential Progression Through the Criminal Justice System: Evidence From Linked Federal and State Court Data
}}
\author{\Large{Brendon McConnell}\thanks{Thanks to Jaime Mill\'{a}n-Quijano, Arnaud Philippe and Mariyana Zapryanova, as well as seminar participants at the CELS2020 conference for comments on an earlier draft. I am grateful to Martin Huber and Giovanni Mellace for sharing code to implement the approach from \citet{Huber2014}, and to J.J. Prescott for permission to use the New Orleans data. Data Disclaimer: Marc L. Miller, J.J. Prescott, and Ronald Wright developed the proprietary NODA data and their documentation from New Orleans District Attorney's Office administrative records, interviews, and institutional research. I was given written permission to use these data for the purpose of this project alone, and I cannot transfer or share them. Questions about these data, their documentation, and their use for research may be directed to Miller, Prescott, or Wright directly. All errors are my own responsibility. Author affiliation and contact: McConnell (Southampton, brendon.mcconnell@gmail.com).}}
\date{\today} 
\begin{document} 
\maketitle
\begin{abstract}
Several key actors -- police, prosecutors, judges -- can alter the course of individuals passing through the multi-staged criminal justice system. I use linked arrest-sentencing data for federal courts from 1994-2010 to examine the role that earlier stages play when estimating Black-white sentencing gaps. I find no evidence of sample selection at play in the federal setting, suggesting federal judges are largely responsible for racial sentencing disparities. In contrast, I document substantial sample selection bias in two different state courts systems. Estimates of racial and ethnic sentencing gaps that ignore selection underestimate the true disparities by 15\% and 13\% respectively. 

\end{abstract}
\vfill

\
\vspace{0mm}
{}

\pagebreak


\newpage
%
\section{\label{sec:Intro}Introduction}
Racial disparities in sentence lengths can stem from the influences of  multiple decision-makers involved in the criminal justice system. This involves the police, who choose whom to arrest, prosecutors, who choose against whom to file charges, as well as the type of charges, and judges, who choose both the type and severity of the sentence. Of those arrested on federal charges, less than one half came before a federal judge for sentencing. Only one in three arrestees come before a state court judge in New Orleans -- one of the state court systems I study in this paper.  Thus, with data on sentencing outcomes, researchers are typically observing at most one half of the original sample of arrestees, raising the possibility of sample selection issues plaguing sentencing regression estimates. 

I use three linked arrest-sentencing datasets for the US criminal justice system (hereafter CJS) and explicitly model the sample selection problem in order to assess the impact of selection bias on the estimated Black-white sentencing gap. For the federal system I construct a data linkage for 1994-2010. For the state system I use both a novel data linkage from 1989-1999 for the New Orleans District Attorney's Office, and linked arrest-sentencing data for state courts in the largest 75 counties.

No prior work has formally accounted for the potential sample selection bias induced by the multi-stage journey defendants take from arrest through to sentencing when analyzing sentencing outcomes. In a prescient piece, \citet{Klepper1983} warned of the hazards of estimating sentencing gaps in the presence of sample selection bias in this setting, carefully spelling out the implications. If there is both (i) \textit{differential stadial progression} by race (i.e., disparate treatment of Black or other minority defendants in the stages preceding sentencing that makes these individuals more likely to end up at the sentencing stage) and (ii) a \textit{sample selection problem} in the sentencing stage (i.e., the unobservables -- such as details of the offense that are known to the prosecutor and judge, but not the econometrician -- that drive stadial progression also impact sentencing outcomes), then this will lead to the Black-white sentencing gap estimates to be biased. In this paper, I consider the following question: to what extent does the estimate of the Black-white sentencing gap change if one accounts for sample selection bias? 

The linked data I use to tackle this topic allows me to follow individuals from arrest, through disposition, to sentencing\footnote{The specifics of case disposition is somewhat different in each of the three cases, and I discuss these difference below.}. For the subset who progress to the sentencing stage, I observe their sentences. I explicitly model progression through the various stages of the CJS using an ordered probit model. This forms the basis of an ordered sample selection model (hereafter an ordered Heckman model), which is an extension of the regular Heckman selection model \citep{Heckman1979}. The flexibility of this ordered selection models allows for both multiple stages in the selection equation, and for there to be an outcome equation associated with more than just one stage. 

To model stadial progression, I define a stage-related variable, $s_i$, as the furthest an individual gets in the CJS. For my baseline approach, this progression begins with (0) an initial arrest stage and moves to (1) filing of charge(s), (2) charging and ends with (3) sentencing\footnote{In order to add context to this approach, consider an example from my main setting -- the federal criminal justice system. An individual is arrested [stage 0]. The assistant US attorney (AUSA) for the district -- the federal prosecutor -- views the case and related evidence and decides to file charges or not. If the case is sufficiently strong, the AUSA files charges and sends the case to a grand jury [stage 1]. The grand jury decide if arrestee will be charged or not [stage 2]. If charged, around 96\% of federal arrestees plead guilty. If the defendant pleads guilty, or is found guilty, they come before a federal district judge for sentencing [stage 3].}. In extensions, I allow for different types of sentencing outcomes.

The exclusion restrictions I use for two of the three empirical settings I consider in this paper are leave-out area-by-year means of the proportion of individuals who are last seen in all except one stage. These leave-out means will capture all relevant systemic factors that impact stadial progression at the area-year level, including the quality of arrest targeting, the nature of the pool of the local population who serve on  grand juries or trial juries, the charging tendencies of the Assistant US Attorney or state prosecutor, and the average sentencing proclivities of the judges serving in the area in the given year. For the remaining setting (New Orleans state courts) I use the leniency of the randomly assigned screening prosecutor as the exclusion restriction. 
Concerns regarding the validity of the exclusion restrictions used for such models are often raised. I tackle these concerns head-on, by testing the validity of the exclusion restrictions I propose, using the methods developed in recent work by \citet{Huber2014}. For all three criminal justice settings that I consider in this paper, there is no sign whatsoever of any violations to the exclusion restrictions I impose -- the $p$-values for the respective tests are equal to 1 in each case. 

Using the universe of U.S. federal sentencing decisions from 1994, I find that Black and white individuals have similar progression from arrest through to sentencing. In addition, I find that there is no significant relationship between the unobservables that drive progression across the stages, and those that determine sentencing. This suggests that federal judges play the predominant role in driving racial disparities in the federal justice system. 

In an extension to my baseline approach for the federal courts, I show that the primary role of judges remains even when I account for mandatory minimum sentence charging -- a key prosecutorial margin highlighted by the work of \citet{RS2014}. Here I separate sentencing into two distinct stages -- sentencing in absence of a mandatory minimum charge, and sentencing under a mandatory minimum. Sentence lengths are considerably longer under mandatory minimum sentences, highlighting the importance of prosecutors on this margin. With this more nuanced approach, I find some evidence of differential stadial progression by race. However, there is still no evidence of sample selection bias in the federal setting.

The absence of a sample selection problem at the sentencing stage means that in the federal system during the time of study, there is no gain to an econometric approach that jointly accounts for selection into sentencing, and the determination of sentence length. It is adequate to estimate sentencing gaps using the sentencing stage in isolation. I return to consider the implications of this null result in Section \ref{sec:SSP}.

In contrast to the federal case, I document differential stadial progression of Black and white individuals in the 1990s New Orleans court system, a period and locale well-known for discriminatory practices that received attention by the U.S. Supreme Court for its level of ``deliberate indifference''\footnote{Justice Ginsburg identified the New Orleans District Attorney's office ``deliberately indifferent'' to the rights of defendants in \textit{Connick v. Thompson}, 563 U.S. 51 (2011).}. Black arrestees are significantly more likely to progress to sentencing than their white counterparts. There is also evidence of sample selection bias -- the unobservables that impact progression through the New Orleans court system are positively and significantly correlated with the unobservables that affect sentence length\footnote{An example of such an unobservable could be a detail of the offense committed e.g., the targeting of a particularly vulnerable victim, or an especially callous action carried out during the offense. If these details are known to the key actors in the CJS -- the screening prosecutor, the sentencing prosecutor, the judge -- but not the econometrician, then such details would be an example of the unobservables that impact both stadial progression, and the sentencing outcome.}. The consequence of this is that the OLS-based Black-white sentencing gap underestimates the true, selection-corrected Black-white sentencing gap by 15\%. 

In the final empirical setting, I expand the scope of state court cases I analyze to consider state courts in the 75 most populous counties in the US\footnote{These counties account for more than a third of the United States population and approximately half of all reported crimes.}. I allow a separate sentencing equation for jail and prison sentences, and in addition the Black-white sentencing gap, I also estimate a Hispanic-white sentencing gap. For Black arrestees, there is no differential stadial progression, so when estimating the Black-white gap, there is no gain from the selection model. There is, however, evidence of significant differential stadial progression for Hispanics. Coupled with a sample selection problem for prison sentencing, this leads to the OLS estimate of the Hispanic-white prison sentencing gap being underestimated by 13\%.

I conclude the paper with a simulation exercise to build on what I learn in this work, and consider the pitfalls of a commonly used alternative approach to dealing with sample selection -- the imputation method or ``Brown'' estimator\footnote{\citet{Chandra2000,Chandra2003} refers to this approach as the ``Brown'' estimator, based on the seminal work by \citet{Brown1984}, who was interested in the related  question of how to correct Black-white earnings ratio when faced with differential non-participation by race.}, whereby one imputes a very low outcome value for all non-participants, and then estimates the relevant parameters using median regression. The implicit assumption here is that sample selection is entirely and overwhelmingly positive, and that it is correct to assign all non-participants an outcome value below the median. One can see examples of this approach in the work of \cite{Neal1996} for the labor market\footnote{A later follow-up piece -- \citet{Johnson2000} -- uses panel data to probe the extremely strong assumption underlying this approach.} and \citet{RS2014} for the criminal justice system. 

Here I make one simple, albeit important, point -- unless one is faced with strong, positive sample selection, it is unwise to simply apply the imputation method in the hope of accounting for selection bias. As my simulations make clear, this approach can, in fact, lead to a \textit{larger bias} than that of ignoring the sample selection problem altogether. This is particularly pertinent in the federal sentencing setting, where I do not find any evidence of a sample selection problem.


Racial disparities in criminal justice has been studied in many fields including economics, law, criminology and sociology. I make a novel contribution to this literature by documenting the consequences of differential stadial progression by race, and sample selection, in a selection model framework that admits the multiple stages of the CJS, and allows for the estimation of sentencing race gaps at different levels (e.g., jail and prison). Prior research has considered the role of decisions-makers or selection at different stages in isolation. One can find several papers that implement a Heckman selection model when considering custodial (court-imposed) vs. non-custodial sentences at the sentencing stage (\citealp{Steffensmeier2001,Ulmer2004}). Other methods include modeling selection with a hazard rate term when analyzing outcomes in a given stage to account for selection from the previous stage alone (\citet{Leiber2003} and \citet{Wooldredge2004}).  \citet{RS2014} show that severity of the prosecutor charge largely renders statistically insignificant racial disparities at the judicial sentencing stage.  Policy-makers have thus focused heightened scrutiny to the role that prosecutors can play in exacerbating racial disparities. None of these papers consider the global progression from arrest to sentence, nor seek to understand the role of sample selection bias.

I make a second novel contribution to the applied sample selection literature, by cautioning against the ``blind'' use of imputation methods -- i.e., using these methods, which assume the strongest form of sample selection bias, without knowing about how much of a sample selection problem exists --  to deal with sample selection concerns.

This paper also adds to the literature that highlights the importance of accounting for sample selection when estimating outcome gaps across protected group status e.g., race and gender gaps in labor market outcomes
\citep{Chandra2000,Mulligan2008,Bar2015}.

\section{\label{sec:DataSetting}Data and Setting}
\subsection{\label{sec:DataSettingFCJS}Federal CJS}
I use linked arrest-sentencing data from US Federal courts spanning the fiscal years 1994-2010 \citep{FJSP}, with the primary focus on the 1994-2003 period. The starting point is determined by data availability. The end point of 2003 was chosen to avoid mixing sentencing outcomes from before and after the \textit{Booker} reforms -- a change in how sentencing guidelines were imposed in the federal system\footnote{See Section \ref{sec:ResultsFCJSsensPostBooker} for more detail on these reforms.}. 

The data appendix in Appendix \ref{sec:AppendixData} details how I link data across the various stages. In selecting my sample, I consider only adult males who are either Black or white\footnote{I do not observe ethnicity at the arrest stage.}. Due to differential treatment under federal law, I remove all non-citizens. Following \citet{RS2014}, who outline several reasons to do so, I omit all immigration arrests. I remove arrest cases for reasons other than a criminal offense (material witness warrants, parole or probation violation) and focus on the 50 US States and the District of Columbia.

Summary statistics are presented in Panel A of Table \ref{tab:global_summstats_1} for my sample of interest. Just under half of Black and white arrestees end up facing a sentencing judge in my data. The $p$-values in column 4 show that there are no significant racial differences at either the start- or the end-point of the federal system. There are, however, large and statistically significant differences at sentencing. Black defendants are 8.7 percentage points more likely to get a custodial sentence, and the average custodial sentence for Black defendants is almost double that of white defendants.

\begin{center}
  \begin{table}[htbp] \centering
\newcolumntype{C}{>{\centering\arraybackslash}X}

\caption{\label{tab:global_summstats_1}Summary Statistics}
{\scriptsize
\begin{tabularx}{\linewidth}{lCCCCCCC}

\toprule
&{(1)}&{(2)}&{(3)}&{(4)}&{(5)}&{(6)}&{(7)} \tabularnewline \midrule
{}&{White}&{Black}&{Hispanic}&{B-W Gap}&{p-Value of B-W Gap}&{H-W Gap}&{p-Value of H-W Gap} \tabularnewline
\midrule \addlinespace[\belowrulesep]
\textbf{A.) Federal Courts}&&&&&&& \tabularnewline
\addlinespace[1ex] Observations&219,889&133,763&&&&& \tabularnewline
\addlinespace[1ex] \textbf{Sentencing Stage}&&&&&&& \tabularnewline
\addlinespace[1ex] 0 -- Arrest&.341&.338&&-.004&[0.882]&& \tabularnewline
1 -- Filing&.114&.120&&.006&[0.057]&& \tabularnewline
2 -- Charging&.062&.053&&-.008&[0.001]&& \tabularnewline
3 -- Sentencing&.483&.489&&.006&[0.819]&& \tabularnewline
\addlinespace[1ex] \textbf{Sentencing Outcomes}&&&&&&& \tabularnewline
\addlinespace[1ex] Custodial Sentence&.831&.918&&.087&[0.000]&& \tabularnewline
\addlinespace[1ex] Sentence Length (Months)&46.4&86.9&&40.5&[0.000]&& \tabularnewline
&(64.8)&(95.6)&&(2.6)&&& \tabularnewline
\addlinespace[1ex] \midrule \addlinespace[2ex] \textbf{B.) New Orleans}&&&&&&& \tabularnewline
\addlinespace[1ex] Observations&21,720&128,254&&&&& \tabularnewline
\addlinespace[1ex] \textbf{Sentencing Stage}&&&&&&& \tabularnewline
\addlinespace[1ex] 0 -- Arrest&.521&.494&&-.027&[0.001]&& \tabularnewline
1 -- Filing&.144&.108&&-.036&[0.000]&& \tabularnewline
2 -- Charging&.073&.054&&-.019&[0.000]&& \tabularnewline
3 -- Sentencing&.262&.344&&.081&[0.000]&& \tabularnewline
\addlinespace[1ex] \textbf{Sentencing Outcomes}&&&&&&& \tabularnewline
\addlinespace[1ex] Custodial Sentence&.941&.976&&.035&[0.000]&& \tabularnewline
\addlinespace[1ex] Sentence Length (Months)&25.7&43.8&&18.1&[0.000]&& \tabularnewline
&(61.6)&(189.9)&&(1.2)&&& \tabularnewline
\addlinespace[1ex] \midrule \addlinespace[2ex] \textbf{C.) Large Urban Counties}&&&&&&& \tabularnewline
\addlinespace[1ex] Observations&28,975&43,917&23,165&&&& \tabularnewline
\addlinespace[1ex] \textbf{Sentencing Stage}&&&&&&& \tabularnewline
\addlinespace[1ex] 0 -- Arrest and Filing&.244&.302&.235&.058&[0.000]&-.009&[0.712] \tabularnewline
1 -- Case Diversion/Deferral&.075&.041&.068&-.033&[0.000]&-.006&[0.416] \tabularnewline
2 -- Charging&.007&.014&.007&.007&[0.001]&-.000&[0.942] \tabularnewline
3 -- Jail Sentencing&.437&.369&.432&-.068&[0.000]&-.005&[0.739] \tabularnewline
4 -- Prison Sentencing&.237&.274&.258&.037&[0.004]&.021&[0.187] \tabularnewline
\addlinespace[1ex] \textbf{Sentencing Outcomes}&&&&&&& \tabularnewline
\addlinespace[1ex] Jail Sentence&.375&.350&.449&-.024&[0.260]&.074&[0.001] \tabularnewline
\addlinespace[1ex] Prison Sentence&.303&.384&.344&.081&[0.000]&.041&[0.003] \tabularnewline
\addlinespace[1ex] Jail Sentence Length (Months)&3.3&3.8&4.1&0.5&[0.032]&0.8&[0.015] \tabularnewline
&(5.3)&(6.5)&(5.2)&(0.2)&&(0.3)& \tabularnewline
Prison Sentence Length (Months)&54.0&61.9&54.7&7.8&[0.030]&0.7&[0.821] \tabularnewline
&(114.2)&(124.1)&(121.9)&(3.5)&&(2.9)& \tabularnewline
\bottomrule \addlinespace[\belowrulesep]

\end{tabularx}
\begin{flushleft}
\scriptsize \textbf{Notes}: Means, standard deviations for continuous variables in parentheses, p-values in square brackets. When testing differences across racial and ethnic groups, standard errors are clustered at a.) district level for the federal CJS data, b.) individual level for New Orleans state courts and c.) county level for the large urban counties state courts data. Given the 2-stage stratified sample design of the data collection for the large urban counties state courts data, weights are used in calculating the summary statistics presented in Panel C.).
\end{flushleft}
}
\end{table}

\end{center}

\subsection{\label{sec:DataSettingNOCJS}New Orleans State CJS}
For New Orleans courts data, my reference period is 1989-1999. I make comparable, though not identical, sample selection decisions when working with this data. My sample of interest is Black and white adult male arrestees. Given that I use information on the screening prosecutor to whom arrestees are assigned, I also require arrestees to face a prosecutor whom I see at least ten times.

Panel B of Table \ref{tab:global_summstats_1} presents summary statistics for my sample of interest. In the New Orleans courts, one can see significant racial differences at each stage of the CJS, immediately contrasting with the federal system. The cumulative effect of this is a large and significant Black-white difference -- 8.1 percentage points or 31\% relative to the white proportion -- in the proportion of arrestees that progress to the final sentencing stage. For those who come before a sentencing judge, Black defendants are more likely to receive a custodial sentence, and receive substantially longer sentence lengths.
\subsection{\label{sec:DataSettingSCPS}75 Largest Counties State CJS}
The final setting I consider are state courts in the 75 most populous counties, with data sourced from the State Courts Processing Statistics series \citep{SCPS}. A two stage-sampling process is implemented in creating this data, hence I use the supplied weights in all analysis. The data are different from the previous two cases along several core dimensions. First, the data identifies ethnicity as well as race, hence I can also estimate Hispanic-white sentencing gaps in this setting. Next, the starting population of interest differs from the other data -- here, felony court filings are used to identify relevant individuals. This means the stage variable looks slightly different (previously I could separately identify individuals who were arrested but has their cases dismissed, and those whose cases progressed to the filing stage.). Finally, the data allows me to identify individuals who face jail and prison sentences. As one can see in Panel C of Table \ref{tab:global_summstats_1}, the associated sentence lengths for the two cases are of a different of magnitude, hence I specify these as two distinct stages. 

\section{\label{sec:EmpSpec}Empirical Specification}
The ordered Heckman approach extends the selection equation of the standard Heckman selection model \citep{Heckman1979} from a probit to an ordered probit. This allows me to jointly estimate i.) the role of race in individuals' progression through the criminal justice system and ii.) racial sentencing disparities in sentencing, conditional on progressing to the sentencing stage.

The first component to the ordered Heckman model is the selection equation, which is where I focus on stadial progression, and define stage, $s_i$, as the furthest an individual advances in the CJS. For my baseline approach, this progression begins with (0) an initial arrest stage and moves to (1) filing of charge(s), (2) charging and ends with (3) sentencing\footnote{
In the sentencing stage, an individual may receive a non-custodial sentence of a fine or probation (which I code as a sentence length of zero). Some papers, focusing solely on the sentencing stage, use a standard Heckman selection model to separately account for the zeroes associated with fines and probation and the positive sentences. I do not want to do that here, but the ordered Heckman approach does admit this approach.}. 
I model this progression with an ordered probit:
\begin{align}
s_{i}^{*} & = X_{i}^{'} \alpha_1 + Z_{s,i}^{'}\alpha_2 + \xi_i \notag \\
          & = Z_{i}^{'}\alpha + \xi_i  \, ;   \notag \\
  s_{i} &=     \begin{cases}
                  \, 0 \quad \text{if } -\infty < s_{i}^{*} \leq \mu_1 \qquad \, \, \, \, \,  \text{[Arrest]}\\
                  \, 1 \quad \text{if } \mu_1 < s_{i}^{*} \leq \mu_2 \qquad \qquad \text{[Filing]} \\
                  \, 2 \quad \text{if } \mu_2 < s_{i}^{*} \leq \mu_3 \qquad \qquad \text{[Charging]} \\
                  \, 3 \quad \text{if } \mu_3  < s_{i}^{*} < \infty  \qquad \qquad \text{[Sentencing]}\, ,
                \end{cases} \label{Eq:oheck_sel}
\end{align}
where $X_i$ is a vector of variables available at the arrest stage -- and thus available for all individuals in the data -- and $Z_{s,i}$ is the exclusion restriction. This vector enters only the selection equation and not the sentencing (or outcome) equation. I detail the specifics of these variables for both the federal and state systems separately below.

The second component to the ordered Heckman model is the sentencing equation:
\begin{equation}
y_{i} =     \begin{cases}
                  X_{i}^{'} \beta + \epsilon_{i} \qquad \, \, \, \text{if } s_{i}=3 \\
                  \text{missing} \qquad \quad \text{otherwise} \, ,  \label{Eq:oheck_out}
            \end{cases} 
\end{equation}
where $y_{i}$, which is only observed if the individual reaches the sentencing stage, is the log of sentence length in months\footnote{In order to account for the zero sentence length observations, I implement a shifted log transform, whereby I add 1 to each sentence length. In order to assess the validity of this, in Sections \ref{sec:ResultsFCJSsensFxlForm} and \ref{sec:ResultsSCJSsens}  I present all core results using an alternative transform -- the inverse hyperbolic sine transformation. The results presented in that section are extremely similar to those of my core specification, which assuages any concerns regarding specific functional form assumptions. That said, I do not provide results for the sentence length in levels -- the extreme right skewness of sentence lengths causes severe issues for the maximum likelihood routine.}, and $\epsilon_{i}$ has mean zero, variance $\sigma^2$ and is bivariate normally distributed with $\xi_i$ with correlation $\rho$. In all of the selection-corrected models that I estimate, I present $\rho$ and a test of whether it is significant or not, as this reflects whether or not there is sample selection bias in the sentencing equation. I estimate equations (\ref{Eq:oheck_sel}) and  (\ref{Eq:oheck_out}) jointly using FIML\footnote{To implement the ordered Heckman routine, I use the user-written Stata package of \citet{CL2007}.}\footnote{The parameter estimates from the FIML procedures are as good as identical to those from the two-step approach, hence I present the more efficient FIML estimates.}.

There is an element of my approach that is somewhat constraining. Given that $X_{i}$ should be a subset of $Z_{i}$ for identification purposes, the vector $X_{i}$ is a set of \textit{arrest-stage} offense type and offender characteristics. It would not be logical to include variables, such as criminal history or presumptive sentence -- commonly used control variables in a sentencing equation -- as these are available \textit{only} for those individuals that I observe at sentencing. The availability of such variables would thus perfectly predict reaching the sentencing stage. The  $X_{i}$ vector is context-specific, and hence described in full in the table notes.

In the federal data, the exclusion restrictions in $Z_{s,i}$ are leave-out district-by-year means of the proportion of individuals who are last seen in stage 0, stage 1 and stage 3\footnote{The combined proportions sum to one, hence I omit stage 2.}. These leave-out means will capture all relevant systemic factors that impact stadial progression at the district-year level, including the quality of arrest targeting, the nature of the pool of the local population who serve on federal grand juries, the charging tendencies of the Assistant US Attorney, and the average sentencing proclivities of the judges serving in the district in the given year\footnote{I can separately identify the parameters associated with the exclusion restrictions -- $\alpha_2$ -- in the presence of district and year fixed effects, as the level of variation for $Z_{s,i}$ is the district-year.}. 

In the New Orleans state court system, a randomly assigned screening prosecutor decides whether or not to progress the case from the arrest stage. I use this as the basis for the exclusion restrictions in $Z_{s,i}$, which are the leave-out screening prosecutor average interacted with a non-missing dummy, and a dummy for missing information on screening prosecutor\footnote{6.6\% of the sample have missing information on screening prosecutor. Instead of just dropping these individuals, I assign them a value of zero for the leave-out screening prosecutor mean, and then create a dummy indicating missing prosecutor information. The results are robust to the alternative approach of merely dropping these observations.}.

For the state courts in large urban counties, the exclusion restrictions in $Z_{s,i}$ are leave-out county-by-year means of the proportion of individuals who are last seen in stage 1-4, with stage 0 serving as the omitted category. The rationale for this choice of exclusion restrictions is the same as for federal courts.

\section{\label{sec:Results}Results}

\subsection{\label{sec:ResultsFCJS}Federal CJS}
Table \ref{tab:oheckman_9403_mylogsentence_SINGLE_5} presents the main set of results based on the federal courts. Column 1 shows the OLS estimate for the raw sentencing gap, whereas column 2 shows the conditional Black-white gap. Moving from column 1 to column 2, I see that the sentencing gap is attenuated to a large degree by the inclusion of a rich set of arrest-level defendant and offense characteristics. The estimated sentencing gap falls from .831 to .349, a 58\% decline. The conditional gap is still extremely large and highly statistically significantly different from zero\footnote{The conditional gap that I document is larger than other papers in the literature. For instance, using a sample that is a subset of the one used here, \citet{MR2021} document an unconditional Black-white sentencing gap of a similar magnitude, but a conditional sentencing gap that is smaller by a factor of 3. The key difference is their paper uses a richer set of covariates, some of which are determined post-arrest, which my empirical strategy precludes.}.

Column 3 shows the coefficient estimate for the Black indicator from the linear index model that underlies the ordered probit. There is no evidence of differential stadial progression by race --the Black coefficient from the selection equation is small and statistically insignificant. This corroborates the lack of racial differences in the unconditional patterns I see in the summary statistics in Panel A of Table \ref{tab:global_summstats_1}.

Column 4 presents the results from an ordered Heckman specification, where equations (\ref{Eq:oheck_sel}) and  (\ref{Eq:oheck_out}) are estimated jointly. First note that when jointly estimating the two equations, I again find no difference in racial progression across the stages. Second, the estimate of $\rho$ -- the correlation between the unobservables that impact stadial progression and sentencing outcomes -- is extremely small, and has a $p$-value of .52. For these two reasons combined, the sentencing gap I estimate using a selection adjusted approach is identical to what I found using a simple OLS. 

This null result is useful in highlighting that both differential stadial progression by race \textit{and} evidence of sample selection bias are required for the ordered selection approach to yield different estimates from a standard OLS. I expand on this point in a simulation exercise in Section \label{sec:MCsims1}.

Finally, the results in column 5 serve as a sensitivity analysis, and confirm my core results. Here I binarize the variable $s_i$, assigning non-sentencing stages a value of zero, and the sentencing stage a value of one. I then implement a standard sample selection model. 

\begin{center}
  \begin{table}[h] \centering
\newcolumntype{C}{>{\centering\arraybackslash}X}

\caption{\label{tab:oheckman_9403_mylogsentence_SINGLE_5} Black-White Sentencing Disparities in the Federal CJS}
{\footnotesize
\begin{tabularx}{\linewidth}{lCCCCC}

\toprule
&{(1)}&{(2)}&{(3)}&{(4)}&{(5)} \tabularnewline \midrule
{}&{OLS}&{OLS}&{Ordered Probit}&{Ordered Heckman}&{Heckman} \tabularnewline
\midrule \addlinespace[\belowrulesep]
\textbf{Sentencing Equation}&&&&& \tabularnewline
\addlinespace[1ex] Black&.831***&.349***&&.349***&.349*** \tabularnewline
&(.0448)&(.0214)&&(.0214)&(.0214) \tabularnewline
\addlinespace[1ex] \textbf{Selection Equation}&&&&& \tabularnewline
\addlinespace[1ex] Black&&&.0116&.0116&.00941 \tabularnewline
&&&(.0114)&(.0114)&(.0107) \tabularnewline
\addlinespace[1ex] \midrule \addlinespace[1ex] Full Set of Controls&&X&X&X&X \tabularnewline
\addlinespace[1ex] \(\overline{sentence}_{W}\)&46.4&46.4&&46.4&46.4 \tabularnewline
\addlinespace[1ex] B-W gap: exp(\(\beta_{Black}\))-1&1.3&.418&.215&.418&.418 \tabularnewline
\addlinespace[1ex] \(\rho\)&&&&.0185&.0107 \tabularnewline
&&&&(.0284)&(.026) \tabularnewline
p-value: \(\rho=0\)&&&&.52&.68 \tabularnewline
p-value: Exclusion Restriction(s)&&&.000&.000&.000 \tabularnewline
\(R^2\)&.0595&.419&&& \tabularnewline
Observations&186,436&186,436&388,123&388,123&388,123 \tabularnewline
\bottomrule \addlinespace[\belowrulesep]

\end{tabularx}
\begin{flushleft}
\scriptsize \textbf{Notes}: *** denotes significance at 1\%, ** at 5\%, and * at 10\%. The dependant variable in the sentencing equation is the log(sentence length in months +1). The +1 is to allow for zero sentence lengths (fines, probation) in the sentencing stage. In the selection equation, the dependant variable is stage, which takes values 0, 1, 2 or 3. Th exception is in specification 6, where I binarize the stage variable (stages 0-2 = 0, stage 3=1). All specifications, with the exception of specification 1, include the following control variables: district dummies, year of arrest dummies, arrest offence code dummies, age decile dummies, marital status dummies and state/country of birth dummies. The exclusion restrictions for the ordered probit selection model are leave-out district\(\times\)year means of the proportion of individuals who are last seen in stage 0, stage 1 and stage 3. For the probit selection model, the exclusion restriction is the leave-out district\(\times\)year mean of the proportion of individuals who are last seen in stage 3. Standard errors are clustered at district level.
\end{flushleft}
}
\end{table}

\end{center}
%
\subsubsection{\label{sec:ResultsFCJSext}Extending the Approach to Account for Mandatory Minimum Charging}
The use of statutory mandatory minimum sentence charging is considered to play a key role in determining the Black-white sentencing gap (\citet{RS2014}).  In this section, I extend my baseline approach outlined in Section \ref{sec:EmpSpec}, in order to incorporate mandatory minimum charging behavior of prosecutors. To do so, I split the final, sentencing stage into two stages -- sentencing absent of a mandatory minimum, and sentencing with a mandatory minimum. I present the results of this extension in Table \ref{tab:oheckman_SM_9403_mylogsentence_SINGLE_5}, and detail this approach in Section \ref{sec:ResultsFCJSextSMM}. The key findings of this extension, is that although this richer model highlights slight differential stadial progression by race, there is still no sample selection problem in either of the sentencing stages, and thus OLS estimates of the sentencing gaps are sufficient here to assess the Black-white sentencing gap.

\subsection{\label{sec:ResultsNOCJS}New Orleans State CJS}
I now move to the  New Orleans state court system, presenting my core results in  Table \ref{tab:oheckman_NODA_mylogsentence_SINGLE_5}. Moving from column 1 to 2, I see that differences in arrest offense and individual characteristics accounts for over half of the raw Black-white sentencing differential, yet a large conditional gap still remains. So far, the analysis follows a similar pattern to the federal case above. 

Where the New Orleans system differs becomes apparent in column 3 -- I see a positive and significant coefficient on Black in the ordered probit equation,  which informs us that the unconditional pattern of differential  stadial progression by race that I document in Panel B of Table \ref{tab:global_summstats_1} persists even when I condition on a rich set of controls. 

Turning to column 4, I note a second point of departure from the federal estimates. The estimate of $\rho$ is large, and highly statistically significantly different from zero, indicating a sample selection problem. That is, the unobservables that determine individuals' progression through the stages of the New Orleans CJS are positively correlated with those that impact sentence severity. Given both the higher likelihood of ending up at the sentencing stages that Black individuals face, and the presence of positive sample selection bias, the coefficients I estimate using the ordered sample selection procedure yield a larger Black-white sentencing gap. Not accounting for differential stadial progression leads to an underestimation of the Black-white sentencing gap by 15\%. 

In the final column I again present a sensitivity analysis of my sample selection approach, simplifying the ordered Heckman model to a standard Heckman by binarizing $s_i$ as I did before. The results tell the same story.

\begin{center}
  \begin{table}[h] \centering
\newcolumntype{C}{>{\centering\arraybackslash}X}

\caption{\label{tab:oheckman_NODA_mylogsentence_SINGLE_5} Black-White Sentencing Disparities in the New Orleans State CJS}
{\footnotesize
\begin{tabularx}{\linewidth}{lCCCCC}

\toprule
&{(1)}&{(2)}&{(3)}&{(4)}&{(5)} \tabularnewline \midrule
{}&{OLS}&{OLS}&{Ordered Probit}&{Ordered Heckman}&{Heckman} \tabularnewline
\midrule \addlinespace[\belowrulesep]
\textbf{Sentencing Equation}&&&&& \tabularnewline
\addlinespace[1ex] Black&.489***&.225***&&.26***&.268*** \tabularnewline
&(.0197)&(.0152)&&(.0157)&(.0158) \tabularnewline
\addlinespace[1ex] \textbf{Selection Equation}&&&&& \tabularnewline
\addlinespace[1ex] Black&&&.133***&.133***&.188*** \tabularnewline
&&&(.0126)&(.0126)&(.0134) \tabularnewline
\addlinespace[1ex] \midrule \addlinespace[1ex] Full Set of Controls&&X&X&X&X \tabularnewline
\addlinespace[1ex] \(\overline{sentence}_{W}\)&25.7&25.7&&25.7&25.7 \tabularnewline
\addlinespace[1ex] B-W gap: exp(\(\beta_{Black}\))-1&.631&.253&&.297&.307 \tabularnewline
\addlinespace[1ex] \(\rho\)&&&&.424&.364 \tabularnewline
&&&&(.0208)&(.0237) \tabularnewline
p-value: \(\rho=0\)&&&&.000&.000 \tabularnewline
p-value: Exclusion Restriction&&&.000&.000&.000 \tabularnewline
\(R^2\)&.0133&.524&&& \tabularnewline
Observations&49,792&49,792&149,970&149,970&149,970 \tabularnewline
\bottomrule \addlinespace[\belowrulesep]

\end{tabularx}
\begin{flushleft}
\scriptsize \textbf{Notes}: *** denotes significance at 1\%, ** at 5\%, and * at 10\%. The dependant variable in the sentencing equation is the log(sentence length in months +1). The +1 is to allow for zero sentence lengths (fines, probation) in the sentencing stage. In the selection equation, the dependant variable is stage, which takes values 0, 1, 2 or 3. Th exception is in specification 6, where we binarize the stage variable (stages 0-2 = 0, stage 3=1). All specifications, with the exception of specification 1, include the following control variables: arresting agency dummies, lead arrest charge dummies, arrest year dummies, age decile dummies, a dummy for multiple arrest charges, and a criminal history dummy. The exclusion restrictions for both the ordered probit, and probit, selection models are the leave-out screening prosecutor mean interacted with a non-missing dummy, and a dummy for missing information on screening prosecutor. Standard errors are clustered at individual level.
\end{flushleft}
}
\end{table}

\end{center}
%
\subsection{\label{sec:ResultsSCJS}State Courts in Large Urban Counties}
I turn now to the final setting I consider in this work -- state courts in the most populous 75 urban counties. Given the nature of the data, I augment my empirical specification slightly to better fit the setting. This amounts to an extension of my baseline ordered Heckman specification, whereby I allow two sentencing stages -- one for jail and another for prison sentences\footnote{I detail the precise nature of this extension in Section \ref{sec:extSCPS}}. 

Table \ref{tab:oheckmanJP_9403_mylogsentence_SINGLE_1} presents the key results for these state courts. The OLS results in columns 1 and 2 highlight the fact that both Black and Hispanic defendants experience a sentencing penalty relative to their white counterparts, for both jail and prison sentences. Tests of equality of parameters cannot be rejected in either setting. Column 3 presents the first indication of a divergence between the two minority groups -- Hispanics experience an additional penalty regarding differential stadial progression, whereas Black individuals are as likely as their white counterparts to progress through the CJS stages.

The selection parameter relevant for jail sentencing -- $\rho_3$ -- is not statistically significantly different from zero, hence there are no differences between the OLS and ordered Heckman parameters for jail sentencing. There is, however, evidence of positive sample selection for prison sentencing in the state court system. The $p$-value for $\hat{\rho}_4$ is .009. Given the combination of differential stadial progression for Hispanic arrestees, and the finding of positive sample selection, I document evidence of a selection-corrected Hispanic-white sentencing gap that is 15\% larger than the non-corrected OLS estimate. Given the lack of differential stadial progression for Black individuals, the selection-corrected Black-white sentencing gap is no different to the one estimated by OLS. Once I account for selection, I find that the Hispanic-white gap is statistically significantly larger than the Black-white gap for prison sentences at conventional levels. 
All key patterns documented here are replicated if I assume a different functional form specification for sentence length. Table \ref{tab:oheckmanJP_9403_myIHSsentence_SINGLE_1} presents the results based on the inverse hyperbolic sine transform.

\begin{center}
  \begin{table}[h] \centering
\newcolumntype{C}{>{\centering\arraybackslash}X}

\caption{\label{tab:oheckmanJP_9403_mylogsentence_SINGLE_1} Black-White Sentencing Disparities in State Courts in Large Urban Counties}
{\footnotesize
\begin{tabularx}{\linewidth}{lCCCCC}

\toprule
&{(1)}&{(2)}&{(3)}&{(4)}&{(5)} \tabularnewline \midrule
\multicolumn{1}{c}{ }& \multicolumn{2}{c}{{OLS}} & \multicolumn{1}{c}{{Ordered}} & \multicolumn{2}{c}{{Ordered}}  \tabularnewline \multicolumn{1}{c}{ }& \multicolumn{2}{c}{} & \multicolumn{1}{c}{{Probit}} & \multicolumn{2}{c}{{Heckman}}  \tabularnewline  \cmidrule(l{2pt}r{5pt}){2-3}  \cmidrule(l{2pt}r{5pt}){4-4}  \cmidrule(l{2pt}r{5pt}){5-6}   \addlinespace[-2ex] \tabularnewline
{}&{Jail}&{Prison}&{}&{Jail}&{Prison} \tabularnewline
\midrule \addlinespace[\belowrulesep]
\textbf{Sentencing Equation}&&&&& \tabularnewline
\addlinespace[1ex] Black&.113***&.0841***&&.113***&.0827*** \tabularnewline
&(.0154)&(.0259)&&(.0151)&(.0264) \tabularnewline
\addlinespace[1ex] Hispanic&.159***&.129***&&.157***&.147*** \tabularnewline
&(.0281)&(.0308)&&(.0306)&(.0332) \tabularnewline
\addlinespace[1ex] \textbf{Selection Equation}&&&&& \tabularnewline
\addlinespace[1ex] Black&&&-.00147&-.0014&-.0014 \tabularnewline
&&&(.0181)&(.018)&(.018) \tabularnewline
\addlinespace[1ex] Hispanic&&&.0419***&.0412***&.0412*** \tabularnewline
&&&(.0135)&(.0135)&(.0135) \tabularnewline
\addlinespace[1ex] \midrule \addlinespace[1ex] \(\overline{sentence}_{W}\)&3.28&53.7&&3.28&53.7 \tabularnewline
\addlinespace[1ex] B-W gap: exp(\(\beta_{Black}\))-1&.12&.0877&&.12&.0863 \tabularnewline
H-W gap: exp(\(\beta_{Hispanic}\))-1&.172&.138&&.17&.159 \tabularnewline
\addlinespace[1ex] \(\rho_3\)&&&&-.057&-.057 \tabularnewline
&&&&(.166)&(.166) \tabularnewline
\(\rho_4\)&&&&.424&.424 \tabularnewline
&&&&(.161)&(.161) \tabularnewline
p-value: \(\rho_3=\rho_4=0\)&&&&.064&.064 \tabularnewline
p-value: Exclusion Restrictions&&&.000&.000&.000 \tabularnewline
\addlinespace[1ex] p-value: \(\beta_{Black} = \beta_{Hispanic}\)&.141&.089&&.187&.023 \tabularnewline
p-value: \(\alpha_{Black} = \alpha_{Hispanic}\)&&&.021&.023&.023 \tabularnewline
\addlinespace[1ex] \(R^2\)&.218&.333&&& \tabularnewline
Observations&38,939&24,945&96,057&96,057&96,057 \tabularnewline
\bottomrule \addlinespace[\belowrulesep]

\end{tabularx}
\begin{flushleft}
\scriptsize \textbf{Notes}: *** denotes significance at 1\%, ** at 5\%, and * at 10\%. Standard errors are clustered at county level. The dependant variable in the sentencing equation is the log(sentence length in months +1). The +1 is to allow for zero sentence lengths (fines, probation) in the sentencing stage. In the selection equation, the dependant variable is stage, which takes values 0, 1, 2, 3 or 4. All specifications include the following control variables: county dummies, year of arrest dummies, most serious arrest offence code dummies, second most serious arrest offence code dummies, dummies for categories of the count of arrest charges, age decile dummies, dummies for most serious prior arrest, prior failure to appear in court, most serious prior conviction and a dummy for prior adult felony conviction for a violent offense. The exclusion restrictions for the ordered probit selection model are leave-out county\(\times\)year means of the proportion of individuals who are last seen in stage 1 and stage 2. Given the 2-stage stratified sample design, SCPS-supplied weights are used in all analysis, thus yielding estimates for the 75 most populous counties in the month of May.
\end{flushleft}
}
\end{table}

\end{center}
%

\section{\label{sec:SSP}The Sample Selection Problem}
\subsection{Testing the Validity of the Exclusion Restrictions}
The sample selection models that I use in this study, although no longer particularly en vogue in the modern applied landscape, are particularly well-suited to studying the multi-stage CJS, where a key outcome of interest occurs only for those who enter the final stage(s). The model enables one to view, in a parsimonious manner, estimates of (i) the differential stadial progression of black arrestees relative to their white counterparts, (ii) the sign and magnitude of the sample selection problem and (iii) the selection-corrected Black-white sentencing gap. 

A key reason that sample selection models are not commonly used is the difficulty in finding a credible exclusion restriction for the selection equation. I formally test the validity of the respective exclusion restrictions I impose in the three empirical settings I study in this work, using the tests proposed by \citet{Huber2014}. In all settings I find overwhelming support for the validity of the exclusion restrictions I impose. The results of these tests can be found in Table \ref{tab:Huber_Mellace_tests_1}, and the test approach described in Section \ref{sec:ExcRestTests}. 
\subsection{A Simulation Exercise}
Here I present a simulation exercise, where I compare the performance of three different estimators -- OLS, the imputation estimator and the Ordered Heckman estimator -- under a variety of different sample selection settings. The aim of this exercise is to gain a better understanding of how the different estimators perform in scenarios that mimic those in the criminal justice system in the US. The exercise highlights the extremely poor performance of the imputation estimator in all scenarios I consider.

The imputation or ``Brown'' estimator is an alternative approach to dealing with sample selection. Here, one imputes a very low outcome value for all non-participants, and then estimates the relevant parameters using median regression. This approach does not require an exclusion restriction or an explicit model for selection. It does however require a very strong assumption -- that selection is so profusely positive, that one can allocate \textit{all} missing outcome values to a sufficiently low value such that all imputed outcomes will be below the median. One can see examples of this approach in the work of \cite{Neal1996} for the labor market and \citet{RS2014} for the criminal justice system. 

I give a brief overview of the simulation approach here, and cover the approach fully in Section \ref{sec:MCsims2}. The nub of this exercise is to simulate a simplified version of equations (\ref{Eq:oheck_sel}) and (\ref{Eq:oheck_out}), and then vary key parameters (specifically those governing (i) the degree of differential stadial progression by race -- $\alpha_1$ below, and (ii) the degree of sample selection bias -- $\rho$).
I simulate data with a sample size of 2,000 -- 1,000 white arrestees and 1,000 Black arrestees. 
I specify the data generating process (DGP) for the selection equation as:
\begin{equation}
s_i^{*} = \alpha_1 Black_i + \alpha_2 Severity_i + \alpha_3 Z_i + \xi_i
\end{equation}
In mapping $s_i^{*}$ to $s_i$, I chose cutoffs to ensure the following proportions in the four stages: 0.20, 0.05, 0.05 and 0.70 for stages 0, 1, 2 and 3 respectively, meaning that 70\% of arrestees will progress to the sentencing stage. 
I specify the DGP for the sentencing equations as:
\begin{equation}
y_{i} =     \begin{cases}
                  \beta_0 + \beta_1 Black_i + \beta_2 Severity_i + \epsilon_{i} \qquad \, \, \, \text{if } s_{i}=3 \\
                  \text{missing} \qquad  \qquad  \qquad  \qquad  \qquad  \qquad \, \, \, \,  \quad \text{otherwise} .  \label{Eq:oheck_out_sim}
            \end{cases} 
\end{equation}
The errors $\xi$ and $\epsilon$ are constructed as standard bivariate normal variables with correlation $\rho$.
For the imputation method, I assign imputed sentences of those 30\% of arrestees who do not make it to the sentencing stage to the lowest sentence observed for those who are sentenced. 

Table \ref{tab:global_simulations_7a_slim} presents the results for three different estimators (OLS, imputation estimators, ordered Heckman\footnote{I cannot estimate the ordered Heckman model for the subset of cases where $\rho=1$, i.e., where selection is perfectly positive.}) of the parameter $\beta_1$ under a variety of different environments defined by (i) the stadial progression and (ii) sample selection parameters. In all cases the true value of $\beta_1 = .10$.
\begin{center}
  \begin{table}[htb] \centering
\newcolumntype{C}{>{\centering\arraybackslash}X}

\caption{\label{tab:global_simulations_7a_slim}Monte Carlo Simulations -- The Imputation Method for Sample Selection Problems}
{\footnotesize
\begin{tabularx}{\linewidth}{lCCCCC}

\toprule
&{(1)}&{(2)}&{(3)}&{(4)}&{(5)} \tabularnewline \midrule
\multicolumn{1}{c}{ }& \multicolumn{2}{c}{{\textbf{Parameter Choices}}} &  \multicolumn{3}{c}{ } \tabularnewline  \cmidrule(l{2pt}r{5pt}){2-3}  \addlinespace[-2ex] \tabularnewline
{}&{\(\bm{\rho}\)}&{\(\bm{\alpha_1}\)}&{\(\bm{\hat{\beta}_1^{OLS}}\)}&{\(\bm{\hat{\beta}_1^{50}}\)}&{\(\bm{\hat{\beta}_1^{OH}}\)} \tabularnewline
\midrule \addlinespace[\belowrulesep]
\addlinespace[2ex] 1.)&1&.50&-.002&.267& \tabularnewline
&&&(.048)&(.083)& \tabularnewline
\addlinespace[.5ex] 2.)&1&.20&.059&.160& \tabularnewline
&&&(.048)&(.080)& \tabularnewline
\addlinespace[.5ex] 3.)&1&.10&.079&.124& \tabularnewline
&&&(.048)&(.080)& \tabularnewline
\addlinespace[.5ex] 4.)&1&0&.100&.089& \tabularnewline
&&&(.048)&(.079)& \tabularnewline
\addlinespace[2ex] 5.)&.50&.50&.049&.365&.100 \tabularnewline
&&&(.053)&(.097)&(.054) \tabularnewline
\addlinespace[.5ex] 6.)&.50&.20&.079&.200&.099 \tabularnewline
&&&(.053)&(.093)&(.053) \tabularnewline
\addlinespace[.5ex] 7.)&.50&.10&.090&.146&.100 \tabularnewline
&&&(.052)&(.092)&(.052) \tabularnewline
\addlinespace[.5ex] 8.)&.50&0&.100&.091&.100 \tabularnewline
&&&(.052)&(.091)&(.053) \tabularnewline
\addlinespace[2ex] 9.)&.25&.50&.074&.398&.100 \tabularnewline
&&&(.053)&(.101)&(.054) \tabularnewline
\addlinespace[.5ex] 10.)&.25&.20&.089&.214&.099 \tabularnewline
&&&(.053)&(.095)&(.054) \tabularnewline
\addlinespace[.5ex] 11.)&.25&.10&.095&.153&.100 \tabularnewline
&&&(.054)&(.094)&(.054) \tabularnewline
\addlinespace[.5ex] 12.)&.25&0&.100&.091&.100 \tabularnewline
&&&(.053)&(.094)&(.053) \tabularnewline
\addlinespace[2ex] 13.)&0&.50&.100&.426&.100 \tabularnewline
&&&(.054)&(.102)&(.055) \tabularnewline
\addlinespace[.5ex] 14.)&0&.20&.100&.225&.100 \tabularnewline
&&&(.054)&(.096)&(.054) \tabularnewline
\addlinespace[.5ex] 15.)&0&.10&.100&.159&.100 \tabularnewline
&&&(.054)&(.095)&(.054) \tabularnewline
\addlinespace[.5ex] 16.)&0&0&.100&.092&.100 \tabularnewline
&&&(.054)&(.096)&(.054) \tabularnewline
\bottomrule \addlinespace[\belowrulesep]

\end{tabularx}
\begin{flushleft}
\scriptsize \textbf{Notes}: Results based on 10,000 simulation runs. The target parameter -- \(\beta_1\) -- is .1 for all simulations. The other fixed parameters are: \(\alpha_2=\alpha_3= \beta_2=1\) and  \( \beta_0=ln(40)\). Column 3 shows the mean and bootrapped standard error of \(\hat{\beta}_1\) from an OLS regression based on the sentencing stage alone. Columns 4 presents the mean and bootrapped standard error of \(\hat{\beta}_1\) from a quantile regression for the 50th percentile based on an imputation approach whereby all missing sentences are allocated the lowest value of sentence length in each iteration run. Column 5 shows the mean and bootrapped standard error of \(\hat{\beta}_1\) from an ordered Heckman regression. Refer to Table \ref{tab:global_simulations_7a} in Section \ref{sec:MCsims2} for a more extensive set of quantile regression results.
\end{flushleft}
}
\end{table}

\end{center}
The core message from this table is that the imputation method is dominated by the ordered Heckman model, but also, in almost all cases, by an OLS approach whereby one ignores any selection concerns. Put differently simply estimating the sentencing gap with OLS without any attempt to correct for sample selection is almost always better than the imputation method.

As one moves down Table \ref{tab:global_simulations_7a_slim}, the sample selection problem -- governed by $\rho$ -- becomes monotonically less of an issue. The final four rows show the estimates of the Black-white sentencing gap when there is no sample selection problem, and is thus a good proxy for the federal courts setting. Focusing on column 4 -- the median regression results -- one can note a severe upwards bias to the imputation based results, particularly for cases when there is differential stadial progression by race. The overestimate is due to the assumption inherent in the imputation approach -- that selection is perfectly positive -- and thus all of those who do not reach the sentencing stage, a group who are disproportionately white (when $\alpha_1>0$), are allocated the lowest observed sentence. Even in the final row, where there is no differential stadial progression and no sample selection problem, the imputation method is still a bad choice, as it gives the wrong standard errors. This is due to the imputed mass point at the far left of the (imputed) sentencing distribution.

These simulation results should caution against the siren song of the seemingly innocuous assumptions of the imputation method.

\section{\label{sec:Conclusion}Conclusion}
Only a sub-sample of individuals who are arrested make it to sentencing. Multiple decision-makers can impact the progression of individuals through the various stages of the criminal justice system. If there are racial disparities in the progression across the CJS stages, then this can bias the measurement of race gaps in sentencing due to sample selection bias.
This paper presents and illustrates a methodology to overcome this bias using using linked arrest-sentencing data for both US federal and state courts systems. My approach offers a way to potentially identify sources of racial disparities.

I conclude the paper with a simulation exercise that will hopefully caution against the use of an alternative approach to account for sample selection at the sentencing stage -- the imputation approach. In this exercise I highlight severe biases to the so-called ``Brown'' estimator, particularly in the setting of mild to no sample selection problem. This particular setting is a good approximation of the federal court system over the entire period of study. The core lesson to take from this exercise is that the imputation approach is globally dominated by the ordered selection model I favor in this paper. Although I show significant biases to using OLS in previous sections, even OLS is preferable to the imputation approach in most scenarios. 

With newly developed methods -- such as those of \citet{Huber2014} -- that allow the practitioner to test the underlying assumptions of the sample selection model, I argue that it is advisable to estimate such models in the sentencing domain. Whether one ultimately will decide to use such models will depend on the context, but it is certainly better to first understand the selection landscape, and then proceed informed.

\clearpage{ }
\bibliography{biblio_CJSstages.bib}

\clearpage{ }
\newpage

\appendix
\section*{\centering{\huge{Appendix}}}
\bigskip

\beginappendixA

\section{\label{sec:AdditionalResults}Additional Results}
\subsection{\label{sec:ResultsFCJSextSMM}Accounting for Sentences With Mandatory Minima -- An Extension}
In this section, I extend the approach I propose in Section \ref{sec:EmpSpec}, in order to incorporate mandatory minimum charging behavior of prosecutors within my framework. To do so, I split the final stage of the ordered probit (see Equation \ref{Eq:oheck_sel}) into two stages -- sentencing absent of a mandatory minimum, and sentencing with a mandatory minimum. 
Average sentence lengths are considerably longer under mandatory minimum sentence, which suggests a natural ordering when assigning values for the extended stage variable.
This results in my estimating conditional sentencing gaps separately based on the presence of a mandatory minimum, and I use exclusion restrictions to aide identification of the factors driving mandatory minimum regime membership.

The modified selection equation can be written as:
\begin{align}
s_{i}^{*} & = X_{i}^{'} \alpha_1 + Z_{s,i}^{'}\alpha_2 + \xi_i \notag \\
          & = Z_{i}^{'}\alpha + \xi_i  \, ;   \notag \\
  s_{i} &=     \begin{cases}
                  \, 0 \quad \text{if } -\infty < s_{i}^{*} \leq \mu_1 \qquad \, \, \, \, \,  \text{[Arrest]}\\
                  \, 1 \quad \text{if } \mu_1 < s_{i}^{*} \leq \mu_2 \qquad \qquad \text{[Filing]} \\
                  \, 2 \quad \text{if } \mu_2 < s_{i}^{*} \leq \mu_3 \qquad \qquad \text{[Charging]} \\
                  \, 3 \quad \text{if } \mu_3  < s_{i}^{*} \leq \mu_4  \qquad \qquad \text{[Sentencing, Mandatory Minimum Absent]}\\
                  \, 4 \quad \text{if } \mu_4  < s_{i}^{*} < \infty  \qquad \qquad \text{[Sentencing, Mandatory Minimum Present]}\, ,
                \end{cases} \label{Eq:oheck_selMM}
\end{align}
where, as before, $X_i$ is a vector of variables available at the arrest stage -- and thus available for all individuals in the data -- and $Z_{s,i}$ is the exclusion restriction. I modify the sentencing component, to allow for sentencing to occur in the two final stages:
\begin{equation}
y_{i} =     \begin{cases}
                  X_{i}^{'} \beta_3 + \epsilon_{i,3} \qquad \, \, \, \text{if } s_{i}=3 \\
                  X_{i}^{'} \beta_4 + \epsilon_{i,4} \qquad \, \, \, \text{if } s_{i}=4 \\
                  \text{missing} \qquad \quad \text{otherwise} \, ,  \label{Eq:oheck_outMM}
            \end{cases} 
\end{equation}

I present the results of my extended ordered Heckman approach in Table \ref{tab:oheckman_SM_9403_mylogsentence_SINGLE_5} below.

\begin{center}
  \begin{table}[h] \centering
\newcolumntype{C}{>{\centering\arraybackslash}X}

\caption{\label{tab:oheckman_SM_9403_mylogsentence_SINGLE_5} Black-White Sentencing Disparities in the Federal CJS}
{\footnotesize
\begin{tabularx}{\linewidth}{lCCCCC}

\toprule
&{(1)}&{(2)}&{(3)}&{(4)}&{(5)} \tabularnewline \midrule
\multicolumn{1}{c}{ }& \multicolumn{2}{c}{{OLS}} & \multicolumn{1}{c}{{Ordered}} & \multicolumn{2}{c}{{Ordered}}  \tabularnewline \multicolumn{1}{c}{ }& \multicolumn{2}{c}{} & \multicolumn{1}{c}{{Probit}} & \multicolumn{2}{c}{{Heckman}}  \tabularnewline  \cmidrule(l{2pt}r{5pt}){2-3}  \cmidrule(l{2pt}r{5pt}){4-4}  \cmidrule(l{2pt}r{5pt}){5-6}   \addlinespace[-2ex] \tabularnewline
{}&{Mandatory Minimum Absent}&{Mandatory Minimum Present}&{}&{Mandatory Minimum Absent}&{Mandatory Minimum Present} \tabularnewline
\midrule \addlinespace[\belowrulesep]
\textbf{Sentencing Equation}&&&&& \tabularnewline
\addlinespace[1ex] Black&.248***&.316***&&.248***&.316*** \tabularnewline
&(.0234)&(.0193)&&(.0243)&(.019) \tabularnewline
\addlinespace[1ex] \textbf{Selection Equation}&&&&& \tabularnewline
\addlinespace[1ex] Black&&&.0549***&.0549***&.0549*** \tabularnewline
&&&(.0106)&(.0106)&(.0106) \tabularnewline
\addlinespace[1ex] \midrule \addlinespace[1ex] \(\rho_3\)&&&&-.0091&-.0091 \tabularnewline
&&&&(.0717)&(.0717) \tabularnewline
\(\rho_4\)&&&&.0044&.0044 \tabularnewline
&&&&(.0508)&(.0508) \tabularnewline
p-value: \(\rho_3=\rho_4=0\)&&&&.98&.98 \tabularnewline
p-value: Exclusion Restriction(s)&&&.000&.000&.000 \tabularnewline
\(R^2\)&.313&.206&&& \tabularnewline
Observations&120,772&65,664&388,123&388,123&388,123 \tabularnewline
\bottomrule \addlinespace[\belowrulesep]

\end{tabularx}
\begin{flushleft}
\scriptsize \textbf{Notes}: *** denotes significance at 1\%, ** at 5\%, and * at 10\%. The dependant variable in the sentencing equation is the log(sentence length in months +1). The +1 is to allow for zero sentence lengths (fines, probation) in the sentencing stage. In the selection equation, the dependant variable is stage, which takes values 0, 1, 2, 3 or 4. All specifications, include the following control variables: district dummies, year of arrest dummies, arrest offence code dummies, age decile dummies, marital status dummies and state/country of birth dummies. The exclusion restrictions for the ordered probit selection model are leave-out district\(\times\)year means of the proportion of individuals who are last seen in stage 0, stage 1 , stage 3 and stage 4. Standard errors are clustered at district level.
\end{flushleft}
}
\end{table}

\end{center}
%

The first point to note, viewing columns 1 and 2, is that the Black-white gap is larger for sentences with a mandatory minimum attached, confirming that prosecutorial charging plays a role in generating racial sentencing differentials.  

Substantial racial sentencing gaps are, however, also present when there are no mandatory minima attached to the case. Some studies suggest that controlling for the prosecutor charge decision, racial sentencing gaps disappear.\footnote{I conducted a set of analyses to relate my findings to those of \citet{RS2014}. There are two key differences that drive the divergence between my respective findings. First, \citet{RS2014} omit offenses relating to drugs, child pornography, traffic offenses and liquor offenses from their main sample. When I do this, my sample size reduces by almost half. This sample selection decision also reduces the Black-white sentencing gap, an effect driven almost entirely by the omission of drug-related offenses. Second,  \citet{RS2014} use a different set of control variables, most notably for the Black-white sentencing differential, the defendant's criminal history. There are significant differences across the races in the criminal history distribution (part of which may reflect disparate racial treatment with previous interactions with the criminal justice system), and controlling for criminal history vastly reduces the Black-white sentencing differential. It is worth noting that my ordered Heckman approach precludes the possibility of controlling for criminal history, which is only available for those who progress to the sentencing stage. Our approach does, however, allow for a correlation between the error terms in the selection and outcome equations. This approach will thus account for the role of criminal history in determining the sub-sample of those who make it to the final sentencing stages.}  The presence of significant racial sentencing gaps for both sets of defendants suggests that federal judges play an important role in racial disparities even when accounting for charges with a mandatory minimum attached.

In a departure from the baseline results in Table \ref{tab:oheckman_9403_mylogsentence_SINGLE_5}, the  Ordered Probit coefficient for Black is now positive and significant, highlighting that Black defendants are more likely to be face a mandatory minimum sentence charge. 

As is the case in my baseline framework, the ordered Heckman makes no difference to the Black-white sentencing gap -- the estimates from columns 4 and 5 are identical to their OLS counterparts in columns 1 and 2. The reason for this is that, even though there is differential stadial progression by race, there is no correlation between the unobservables that affect progression and those that drive sentence length -- $\rho_4$ and  $\rho_5$ are extremely small, and the $p$-value of a joint test of $\rho_4$ and  $\rho_5$ is 0.98. This result highlights again that in order for the ordered Heckman approach to deliver estimates that differ from a simple OLS, there needs to be differential racial progression through the CJS \textit{and} a significant correlation between the unobservables influencing progression and those influencing sentencing outcomes.

\subsection{\label{sec:extSCPS}Allowing for two Sentencing Levels -- An Extension}
The nature of the large urban county state court data differs somewhat from the other two cases. The data tracks cases from arrest to sentencing for individuals against whom a felony case has been filed. This is already a point of departure, as in the other settings I specify arrest and case filing as two distinct stages. Next, individuals can receive a jail or a prison sentence. I allow for these two distinct outcomes to be two separate stages. Finally, the details available to me regarding case disposition enable me to disambiguate between different intermediate outcomes. Specifically I can separately identify defendants whose cases where diverted or deferred\footnote{The use of the option of diversion and deferment has increased over time. In the state court system in large urban counties in the 1990s, the option was used in 4.1\% of cases, whereas in the next decade it was used in 7.4\% of cases. \citet{MuellerSmith2020} study the consequences of diversion, documenting large, beneficial impacts on the margins of both crime and the labor market for those who receive such outcomes.}, from those who were formally charged. Diversion. Given these differences, it is worth rewriting the order Heckman specification for this setting.
\begin{align}
s_{i}^{*} & = X_{i}^{'} \alpha_1 + Z_{s,i}^{'}\alpha_2 + \xi_i \notag \\
          & = Z_{i}^{'}\alpha + \xi_i  \, ;   \notag \\
  s_{i} &=     \begin{cases}
                  \, 0 \quad \text{if } -\infty < s_{i}^{*} \leq \mu_1 \qquad \, \, \, \, \,  \text{[Arrest and Filing]}\\
                  \, 1 \quad \text{if } \mu_1 < s_{i}^{*} \leq \mu_2 \qquad \qquad \text{[Case Diverted/Deferred]} \\
                  \, 2 \quad \text{if } \mu_2 < s_{i}^{*} \leq \mu_3 \qquad \qquad \text{[Charging]} \\
                  \, 3 \quad \text{if } \mu_3  < s_{i}^{*} \leq \mu_4  \qquad \qquad \text{[Jail Sentencing]}\, \\
                  \, 4 \quad \text{if } \mu_4  < s_{i}^{*} < \infty  \qquad \qquad \text{[Prison Sentencing]}\, ,
                \end{cases} \label{Eq:oheck_selSCPS}
\end{align}
where $X_i$ is a vector of variables available at the arrest stage -- and thus available for all individuals in the data -- and $Z_{s,i}$ is the exclusion restriction. This vector enters only the selection equation and not the sentencing (or outcome) equation. The second component to the ordered Heckman model is the sentencing equation:
\begin{equation}
y_{i} =     \begin{cases}
                  X_{i}^{'} \beta_3 + \epsilon_{i,3} \qquad \, \, \, \text{if } s_{i}=3 \\
                  X_{i}^{'} \beta_4 + \epsilon_{i,4} \qquad \, \, \, \text{if } s_{i}=4 \\
                  \text{missing} \qquad \quad \text{otherwise} \, ,  \label{Eq:oheck_outSCPS}
            \end{cases} 
\end{equation}
\newpage{ }
\subsection{\label{sec:ExcRestTests}Testing the Exclusion Restrictions}
I make two restrictions when testing the exclusion restrictions, both to fit the presentation of the testing procedure developed in \citet{Huber2014}. First, I collapse my multi-stage (ordered probit) approach to specifying the selection equation into a binary (standard probit) one -- as I do in column 5 of both Tables \ref{tab:oheckman_9403_mylogsentence_SINGLE_5} and \ref{tab:oheckman_NODA_mylogsentence_SINGLE_5}. Next, the authors present the case for a binary exclusion restriction. In both their empirical applications, the authors start with a non-binary exclusion restriction (husbands income, number of children), and then binarize these. I follow the authors, and binarize my instrument.

It is beyond the scope of this appendix section to cover the entire approach of \citet{Huber2014} -- I encourage the interested reader to consult the paper directly. In this section, my aim is to explain what the results in Table \ref{tab:Huber_Mellace_tests_1} are telling us.

Although I focus on the exclusion restriction, the approach of \citet{Huber2014} is in fact testing both the validity of the exclusion restriction \textit{and} the additive separability of the error term in the selection equation. The authors show that this second test can be represented as a test of monotonicity of the selection state with respect to the exclusion restriction, where monotonicity can be positive (as in my case) or negative. 

The authors then work through a series of calculations to arrive at two bounding rules for the specific sub-group of individuals who would make it to the sentencing stage irrespective of what realization of the exclusion restriction, $Z_{s,i}$, that they receive (the always-takers) -- one for probability measures of the outcome distribution, and another for the mean outcome. These bounding rules give rise to the two sets of inequality constraints that form the basis of the mean-based and probability-based tests, the $p$-values of which are presented in Table \ref{tab:Huber_Mellace_tests_1}. Finally, the standardized difference presented in the first row of this table represents the magnitude of the maximum constraint violation for the mean-based test -- the $p$-value in row 2 represents the precision of any such violation. As the authors note, a negative standardized difference indicates that no constraint is violated.

\begin{center}
  \begin{table}[htb] \centering
\newcolumntype{C}{>{\centering\arraybackslash}X}

\caption{\label{tab:Huber_Mellace_tests_1}Testing the Exclusion Restrictions -- \citet{Huber2014}}
{\footnotesize
\begin{tabularx}{\linewidth}{lCCC}

\toprule
&{(1)}&{(2)}&{(3)} \tabularnewline \midrule
{}&{Federal}&{New Orleans}&{75 Largest Urban Counties} \tabularnewline
\midrule \addlinespace[\belowrulesep]
\addlinespace[.5ex]Standardized Difference&--0.372&--0.288&--0.329 \tabularnewline
\addlinespace[.5ex]p-Value Mean-Based Constraints&1.000&1.000&1.000 \tabularnewline
\addlinespace[.5ex]p-Value Probability-Based Constraints&1.000&1.000&1.000 \tabularnewline
\bottomrule \addlinespace[\belowrulesep]

\end{tabularx}
\begin{flushleft}
\scriptsize \textbf{Notes}: Results based on 10,000 simulation runs. The first row -- Standardized Difference -- refers to standardized maximum of the mean constraints in Equation 12 of \citet{Huber2014}, who note that a negative or zero value implies that no constraint is violated. The next two rows are based on test of the mean-based (Equation 12 of  \citet{Huber2014}) and probability-based constraints (Equation 8 of \citet{Huber2014}) respectively.
\end{flushleft}
}
\end{table}

\end{center}

\clearpage{ }

\subsection{Differential Arrest by Race}
This work considers the consequence of differential stadial progression by race, taking the arrest stage as the starting point. This decision is entirely data-driven.
A large body of work documents racial bias in police behavior during in civilian interactions prior to and at arrest \citep{Antonovics2009,Pierson2020,Ba2021,Hoekstra2022}.
If there is racial bias in arrest risk, then the approach taken here will miss this important stage. Put differently, my arrest sample will be a selected sample of those who commit a crime.

Hypothetically, if there were a data linkage from the suspicion of crime phase through to arrest and then sentencing, then the framework presented here could easily accommodate this extension, e.g., by specifying a pre-arrest phase -- stage -1. In this setting, if Black individuals suspected of a crime were more likely to be arrested than their white counterparts ($\alpha_{Black}>0$), and if there are details of the suspected crime that will later impact sentencing and  are observed by police but not the econometrician ($\rho>0$), then even the selection-corrected estimates of the Black-white sentencing gap presented here will be an underestimate of the true sentencing gap, one that incorporates racial disparities in the arrest decision. 

In this section I create two sets of Black-white and Hispanic-white ratios -- for the population relevant to each courts system, and for arrestees. Columns 3 and 6 are the arrest ratio to population ratio. A ratio of ratios equal to one would indication equal, proportional representation at arrest across racial and ethnic groups. A ratio of ratios above 1 signals over-representation for the minority group in question, below 1 the converse.

Table \ref{tab:ratios} documents the ratios of Black:white and Hispanic:white arrestee ratios in both the data employed here, and the relevant source populations of these arrestees. The latter is not precisely what I want (which are the race and ethnic ratios of those suspected of committing a crime), but is as close as I can get. What the statistics in this table show is that Black and Hispanic individuals are over-represented in the pool of arrestees -- the starting point for my analysis -- by a factor that ranges from 2 to over 5.5. Unless the sole reason for this vast difference in representation at arrest is due to differences in committing crime across groups, then this suggests my selection-corrected sentencing gap estimates are lower bounds of the true racial and ethnic disparities in sentencing. 
\begin{center}
  \begin{table}[h] \centering
\newcolumntype{C}{>{\centering\arraybackslash}X}

\caption{\label{tab:ratios} Comparing Population and Arrestee Race and Ethnicity Ratios}
{\footnotesize
\begin{tabularx}{\linewidth}{lCCCCCC}

\toprule
&{(1)}&{(2)}&{(3)}&{(4)}&{(5)}&{(6)} \tabularnewline \midrule
\multicolumn{1}{c}{ }& \multicolumn{3}{c}{{Black:White}} & \multicolumn{3}{c}{{{Hispanic:White}}}  \tabularnewline  \cmidrule(l{2pt}r{5pt}){2-4}  \cmidrule(l{2pt}r{5pt}){5-7}  \tabularnewline
{}&{Population Ratio}&{Arrestee  Ratio}&{Arrestee Ratio: Population  Ratio}&{Population Ratio}&{Arrestees Ratio}&{Arrestee Ratio: Population  Ratio} \tabularnewline
\midrule \addlinespace[\belowrulesep]
\addlinespace[1ex] \textbf{Federal Courts}&.142&.608&4.30&&& \tabularnewline
\addlinespace[1ex] \textbf{New Orleans}&2.40&5.90&2.46&&& \tabularnewline
\addlinespace[1ex] \textbf{Largest 75 Counties}&.290&1.63&5.62&.397&.844&2.13 \tabularnewline
\bottomrule \addlinespace[\belowrulesep]

\end{tabularx}
\begin{flushleft}
\scriptsize \textbf{Notes}:The sources for the arrestee ratios are the three datasets I used in the paper -- the FJSP data for the federal system, the NODA data I use for the New Orleans state system and the SCPS data I use for the large urban counties state courts system. To calculate adult male population ratios for the federal courts, I use data from the 2000 Census, specifically 2000 Census Table 5 from \url{https://www.census.gov/data/tables/2000/dec/phc-t-09.html}. For New Orleans, I use statistics for the New Orleans city population in 2000 from \url{http://censusviewer.com/city/LA/New\%20Orleans}. Finally, for the 75 largest urban counties, I select the 75 counties from the Census 2000 county-level data here -- \url{https://www.census.gov/data/datasets/time-series/demo/popest/intercensal-2000-2010-counties.html}.
\end{flushleft}
}
\end{table}

\end{center}
\newpage{}
\subsection{\label{sec:ResultsFCJSsens}Federal CJS Sensitivity Analyses}

I probe the baseline federal results in several directions, in order to assess (i) the validity of the (shift log transform) functional form assumption, (ii) the stability of the core, pre-\textit{Booker} results in the post-\textit{Booker} period and (iii) whether it is the granularity of the arrest offense controls that are driving the finding of no sample selection.
\subsubsection{\label{sec:ResultsFCJSsensFxlForm}Functional Form}
Table \ref{tab:oheckman_9403_myIHSsentence_SINGLE_5} presents the results for my baseline setting, but where the dependent variable in the outcome equation is the inverse hyperbolic sine of sentence length instead of the shifted log transform. This alternative transformation is becoming more popular in criminal justice papers, for example, \citet{Feigenberg2021}, \citet{Norris2021}and  \citet{Williams2022}.

 Like the natural log, the IHS is a concave transformation, and thus deals with the extreme (right) skewness of the sentencing data. The results presented below confirm what I find {in the main analysis (using a log specification), both qualitatively and quantitatively.
\begin{center}
  \begin{table}[h] \centering
\newcolumntype{C}{>{\centering\arraybackslash}X}

\caption{\label{tab:oheckman_9403_myIHSsentence_SINGLE_5}  Functional Form Sensitivity Analysis -- Federal CJS}
{\footnotesize
\begin{tabularx}{\linewidth}{lCCCCC}

\toprule
&{(1)}&{(2)}&{(3)}&{(4)}&{(5)} \tabularnewline \midrule
{}&{OLS}&{OLS}&{Ordered Probit}&{Ordered Heckman}&{Heckman} \tabularnewline
\midrule \addlinespace[\belowrulesep]
\textbf{Sentencing Equation}&&&&& \tabularnewline
\addlinespace[1ex] Black&.908***&.379***&&.379***&.379*** \tabularnewline
&(.0501)&(.0242)&&(.0241)&(.0241) \tabularnewline
\addlinespace[1ex] \textbf{Selection Equation}&&&&& \tabularnewline
\addlinespace[1ex] Black&&&.0116&.0116&.00941 \tabularnewline
&&&(.0114)&(.0114)&(.0107) \tabularnewline
\addlinespace[1ex] \midrule \addlinespace[1ex] Full Set of Controls&&X&X&X&X \tabularnewline
\addlinespace[1ex] \(\overline{sentence}_{W}\)&46.4&46.4&&46.4&46.4 \tabularnewline
\addlinespace[1ex] \(\rho\)&&&&.0161&.00899 \tabularnewline
&&&&(.0258)&(.0238) \tabularnewline
p-value: \(\rho=0\)&&&&.53&.71 \tabularnewline
p-value: Exclusion Restriction(s)&&&.000&.000&.000 \tabularnewline
\(R^2\)&.0557&.406&&& \tabularnewline
Observations&186,436&186,436&388,123&388,123&388,123 \tabularnewline
\bottomrule \addlinespace[\belowrulesep]

\end{tabularx}
\begin{flushleft}
\scriptsize \textbf{Notes}: *** denotes significance at 1\%, ** at 5\%, and * at 10\%. The dependant variable in the sentencing equation is the inverse hyperbolic sine of sentence length in months. In the selection equation, the dependant variable is stage, which takes values 0, 1, 2 or 3. Th exception is in specification 6, where I binarize the stage variable (stages 0-2 = 0, stage 3=1). All specifications, with the exception of specification 1, include the following control variables: district dummies, year of arrest dummies, arrest offence code dummies, age decile dummies, marital status dummies and state/country of birth dummies. The exclusion restrictions for the ordered probit selection model are leave-out district\(\times\)year means of the proportion of individuals who are last seen in stage 0, stage 1 and stage 3. For the probit selection model, the exclusion restriction is the leave-out district\(\times\)year mean of the proportion of individuals who are last seen in stage 3. Standard errors are clustered at district level.
\end{flushleft}
}
\end{table}

\end{center}
\subsubsection{\label{sec:ResultsFCJSsensPostBooker}The Post-\textit{Booker} Period}
The Supreme Court decision in \textit{U.S. v. Booker and Fanfan} -- that the previously mandatory federal sentencing guidelines should now hereafter be considered merely in an advisory capacity -- strongly suggests that I should consider the pre-\textit{Booker} and post-\textit{Booker} periods as two separate regimes.

The reason I truncate the data at the end of the 2003 fiscal year is that during the 2004 fiscal year, the Washington state supreme court decided on \textit{Blakely v. Washington}, an antecedent to \textit{U.S. v. Booker and Fanfan}, which led to certain judges deciding they could not apply the sentencing guidelines, either fully or partially, on constitutional grounds. Some did so on a case-by-case basis, whilst others followed the local norm established with their district. This was the case even though these were federal judges hearing federal cases, and \textit{Blakely} was a state case. For this reason, I omit the fiscal years of 2004 and 2005 from analysis, and instead focus on the two homogeneous periods of i.) Pre-\textit{Blakely} -- 1994-2003 and ii.) Post-\textit{Booker} -- 2006 onward.

In this section I replicate my main analysis (which focused on the pre-\textit{Blakeley} period of 1994-2003), and consider the post-\textit{Booker} period of 2006-2010. Table \ref{tab:oheckman_0610_mylogsentence_SINGLE_5} repeats my core analysis, but for the post-\textit{Booker} years. Table \ref{tab:oheckman_0610_myIHSsentence_SINGLE_5} probes the sensitivity of these findings to the functional form specification, using the inverse hyperbolic sine in place of the log specification. The two points to take from these two tables are i.) the post-\textit{Booker} results are qualitatively identical to what I find in my core analysis and ii.) the results are not sensitive to the functional form specification.

\begin{center}
  \begin{table}[h] \centering
\newcolumntype{C}{>{\centering\arraybackslash}X}

\caption{\label{tab:oheckman_0610_mylogsentence_SINGLE_5} Black-White Sentencing Disparities in the Federal CJS -- 2006-2010}
{\footnotesize
\begin{tabularx}{\linewidth}{lCCCCC}

\toprule
&{(1)}&{(2)}&{(3)}&{(4)}&{(5)} \tabularnewline \midrule
{}&{OLS}&{OLS}&{Ordered Probit}&{Ordered Heckman}&{Heckman} \tabularnewline
\midrule \addlinespace[\belowrulesep]
\textbf{Sentencing Equation}&&&&& \tabularnewline
\addlinespace[1ex] Black&.703***&.376***&&.376***&.376*** \tabularnewline
&(.0504)&(.0208)&&(.0208)&(.0208) \tabularnewline
\addlinespace[1ex] \textbf{Selection Equation}&&&&& \tabularnewline
\addlinespace[1ex] Black&&&.014&.014&.00647 \tabularnewline
&&&(.0182)&(.0182)&(.0157) \tabularnewline
\addlinespace[1ex] \midrule \addlinespace[1ex] Full Set of Controls&&X&X&X&X \tabularnewline
\addlinespace[1ex] \(\rho\)&&&&.00456&-.00361 \tabularnewline
p-value: \(\rho=0\)&&&&.76&.83 \tabularnewline
p-value: Exclusion Restriction(s)&&&.000&.000&.000 \tabularnewline
\(R^2\)&.048&.372&&& \tabularnewline
Observations&85,311&85,311&215,722&215,722&215,722 \tabularnewline
\bottomrule \addlinespace[\belowrulesep]

\end{tabularx}
\begin{flushleft}
\scriptsize \textbf{Notes}: *** denotes significance at 1\%, ** at 5\%, and * at 10\%. The dependant variable in the sentencing equation is the log(sentence length in months +1). The +1 is to allow for zero sentence lengths (fines, probation) in the sentencing stage. In the selection equation, the dependant variable is stage, which takes values 0, 1, 2 or 3. Th exception is in specification 6, where we binarize the stage variable (stages 0-2 = 0, stage 3=1). All specifications, with the exception of specification 1, include the following control variables: district dummies, year of arrest dummies, arrest offence code dummies, age decile dummies, marital status dummies and state/country of birth dummies. The exclusion restrictions for the ordered probit selection model are leave-out district\(\times\)year means of the proportion of individuals who are last seen in stage 0, stage 1 and stage 3. For the probit selection model, the exclusion restriction is the leave-out district\(\times\)year mean of the proportion of individuals who are last seen in stage 3. Standard errors are clustered at district level.
\end{flushleft}
}
\end{table}

  \begin{table}[h] \centering
\newcolumntype{C}{>{\centering\arraybackslash}X}

\caption{\label{tab:oheckman_0610_myIHSsentence_SINGLE_5} Functional Form Sensitivity Analysis -- Federal CJS -- 2006-2010}
{\footnotesize
\begin{tabularx}{\linewidth}{lCCCCC}

\toprule
&{(1)}&{(2)}&{(3)}&{(4)}&{(5)} \tabularnewline \midrule
{}&{OLS}&{OLS}&{Ordered Probit}&{Ordered Heckman}&{Heckman} \tabularnewline
\midrule \addlinespace[\belowrulesep]
\textbf{Sentencing Equation}&&&&& \tabularnewline
\addlinespace[1ex] Black&.767***&.412***&&.412***&.412*** \tabularnewline
&(.0529)&(.0233)&&(.0232)&(.0233) \tabularnewline
\addlinespace[1ex] \textbf{Selection Equation}&&&&& \tabularnewline
\addlinespace[1ex] Black&&&.014&.014&.00647 \tabularnewline
&&&(.0182)&(.0182)&(.0157) \tabularnewline
\addlinespace[1ex] \midrule \addlinespace[1ex] Full Set of Controls&&X&X&X&X \tabularnewline
\addlinespace[1ex] \(\rho\)&&&&.00333&-.00436 \tabularnewline
p-value: \(\rho=0\)&&&&.81&.77 \tabularnewline
p-value: Exclusion Restriction(s)&&&.000&.000&.000 \tabularnewline
\(R^2\)&.0454&.361&&& \tabularnewline
Observations&85,311&85,311&215,722&215,722&215,722 \tabularnewline
\bottomrule \addlinespace[\belowrulesep]

\end{tabularx}
\begin{flushleft}
\scriptsize \textbf{Notes}: *** denotes significance at 1\%, ** at 5\%, and * at 10\%. The dependant variable in the sentencing equation is the inverse hyperbolic sine of sentence length in months. In the selection equation, the dependant variable is stage, which takes values 0, 1, 2 or 3. Th exception is in specification 6, where we binarize the stage variable (stages 0-2 = 0, stage 3=1). All specifications, with the exception of specification 1, include the following control variables: district dummies, year of arrest dummies, arrest offence code dummies, age decile dummies, marital status dummies and state/country of birth dummies. The exclusion restrictions for the ordered probit selection model are leave-out district\(\times\)year means of the proportion of individuals who are last seen in stage 0, stage 1 and stage 3. For the probit selection model, the exclusion restriction is the leave-out district\(\times\)year mean of the proportion of individuals who are last seen in stage 3. Standard errors are clustered at district level.
\end{flushleft}
}
\end{table}

\end{center}
\,
\clearpage{ }
\,
\subsubsection{\label{sec:ResultsFCJSsensXgranularity}Is the Granular Detail of the Control Variables Driving the Federal Results?}
Table \ref{tab:oheckman_9403_mylogsentence_SINGLE_5a} presents results for three different levels of arrest offense controls. The ``coarse'' controls account for 8 main arrest categories, the ``medium'' controls account for 55 arrest sub-categories and the ``fine'' controls -- also my baseline controls -- account for 312 distinct arrest offenses. Unsurprisingly the Black-white sentencing gap differs across the specifications, but the key message from this set of results is that the granularity of my baseline controls is not driving the sample selection parameter estimates. Even with very coarse offense controls, I still find no evidence of a sample selection problem, no differential stadial progression by race, and thus no difference between the OLS and ordered Heckman estimates of the Black-white sentencing gap.

\begin{center}
  \begin{table}[h] \centering
\newcolumntype{C}{>{\centering\arraybackslash}X}

\caption{\label{tab:oheckman_9403_mylogsentence_SINGLE_5a} Black-White Sentencing Disparities in the Federal CJS}
{\footnotesize
\begin{tabularx}{\linewidth}{lCCCCCC}

\toprule
&{(1)}&{(2)}&{(3)}&{(4)}&{(5)}&{(6)} \tabularnewline
\multicolumn{1}{c}{ }& \multicolumn{3}{c}{{OLS}} & \multicolumn{3}{c}{{Ordered Heckman}}  \tabularnewline  \cmidrule(l{2pt}r{5pt}){2-4}  \cmidrule(l{2pt}r{5pt}){5-7}   \addlinespace[-2ex] \tabularnewline
\midrule \addlinespace[\belowrulesep]
\textbf{Sentencing Equation}&&&&&& \tabularnewline
\addlinespace[1ex] Black&.454***&.361***&.349***&.454***&.361***&.349*** \tabularnewline
&(.0307)&(.0225)&(.0214)&(.0306)&(.0225)&(.0214) \tabularnewline
\addlinespace[1ex] \textbf{Selection Equation}&&&&&& \tabularnewline
\addlinespace[1ex] Black&&&&.016&.0105&.0116 \tabularnewline
&&&&(.0171)&(.0123)&(.0114) \tabularnewline
\addlinespace[1ex] \midrule \addlinespace[1ex] Offense Control Type&Coarse&Medium&Fine&Coarse&Medium&Fine \tabularnewline
\addlinespace[1ex] \(\rho\)&&&&.0253&.0201&.0185 \tabularnewline
&&&&(.0309)&(.0291)&(.0284) \tabularnewline
p-value: \(\rho=0\)&&&&.41&.49&.52 \tabularnewline
yvarWhite&46.4&46.4&46.4&46.4&46.4&46.4 \tabularnewline
expBetaBlack&.575&.435&.418&.575&.435&.418 \tabularnewline
p-value: Exclusion Restriction(s)&&&&.000&.000&.000 \tabularnewline
\(R^2\)&.359&.406&.419&&& \tabularnewline
Observations&186,436&186,436&186,436&388,123&388,123&388,123 \tabularnewline
\bottomrule \addlinespace[\belowrulesep]

\end{tabularx}
\begin{flushleft}
\scriptsize \textbf{Notes}: *** denotes significance at 1\%, ** at 5\%, and * at 10\%. The dependant variable in the sentencing equation is the log(sentence length in months +1). The +1 is to allow for zero sentence lengths (fines, probation) in the sentencing stage. In the selection equation, the dependant variable is stage, which takes values 0, 1, 2 or 3. Th exception is in specification 6, where I binarize the stage variable (stages 0-2 = 0, stage 3=1). All specifications include the following control variables: district dummies, year of arrest dummies, age decile dummies, marital status dummies and state/country of birth dummies. The coarse controls are a series of dummies for the 8 main offense categories. The medium controls are a series of dummies for the 55 sub-offense categories. The fine controls are the baseline offense controls -- a series of dummies for the 312 arrest offenses. The exclusion restrictions for the ordered probit selection model are leave-out district\(\times\)year means of the proportion of individuals who are last seen in stage 0, stage 1 and stage 3. Standard errors are clustered at district level.
\end{flushleft}
}
\end{table}

\end{center}
\newpage{}
\subsection{\label{sec:ResultsSCJSsens}State CJS Sensitivity Analyses}
Here I present the results for the two state CJS settings I consider in this work, using the inverse hyperbolic sine (IHS) transformation of sentence length instead of shifted log transform. In both cases, the core patterns documented in the main text are replicated using this alternative approach to deal with the extreme (right) skewness of the sentencing data. 
\begin{center}
  \begin{table}[htb] \centering
\newcolumntype{C}{>{\centering\arraybackslash}X}

\caption{\label{tab:oheckman_NODA_myIHSsentence_SINGLE_5}  Functional Form Sensitivity Analysis -- New Orleans State CJS}
{\footnotesize
\begin{tabularx}{\linewidth}{lCCCCC}

\toprule
&{(1)}&{(2)}&{(3)}&{(4)}&{(5)} \tabularnewline \midrule
{}&{OLS}&{OLS}&{Ordered Probit}&{Ordered Heckman}&{Heckman} \tabularnewline
\midrule \addlinespace[\belowrulesep]
\textbf{Sentencing Equation}&&&&& \tabularnewline
\addlinespace[1ex] Black&.547***&.255***&&.284***&.291*** \tabularnewline
&(.0223)&(.0174)&&(.0177)&(.0177) \tabularnewline
\addlinespace[1ex] \textbf{Selection Equation}&&&&& \tabularnewline
\addlinespace[1ex] Black&&&.133***&.133***&.188*** \tabularnewline
&&&(.0126)&(.0126)&(.0134) \tabularnewline
\addlinespace[1ex] \midrule \addlinespace[1ex] Full Set of Controls&&X&X&X&X \tabularnewline
\addlinespace[1ex] \(\overline{sentence}_{W}\)&25.7&25.7&&25.7&25.7 \tabularnewline
\addlinespace[1ex] \(\rho\)&&&&.325&.278 \tabularnewline
&&&&(.0181)&(.0181) \tabularnewline
p-value: \(\rho=0\)&&&&.000&.000 \tabularnewline
p-value: Exclusion Restriction&&&.000&.000&.000 \tabularnewline
\(R^2\)&.0135&.511&&& \tabularnewline
Observations&49,792&49,792&149,970&149,970&149,970 \tabularnewline
\bottomrule \addlinespace[\belowrulesep]

\end{tabularx}
\begin{flushleft}
\scriptsize \textbf{Notes}: *** denotes significance at 1\%, ** at 5\%, and * at 10\%. The dependant variable in the sentencing equation is the inverse hyperbolic since of sentence length in months. In the selection equation, the dependant variable is stage, which takes values 0, 1, 2 or 3. Th exception is in specification 6, where we binarize the stage variable (stages 0-2 = 0, stage 3=1). All specifications, with the exception of specification 1, include the following control variables: arresting agency dummies, lead arrest charge dummies, arrest year dummies, age decile dummies, a dummy for multiple arrest charges, and a criminal history dummy. The exclusion restrictions for both the ordered probit, and probit, selection models are the leave-out screening prosecutor mean interacted with a non-missing dummy, and a dummy for missing information on screening prosecutor. Standard errors are clustered at individual level.
\end{flushleft}
}
\end{table}

\end{center}
\vspace{-1pt}
\begin{center}
  \begin{table}[h] \centering
\newcolumntype{C}{>{\centering\arraybackslash}X}

\caption{\label{tab:oheckmanJP_9403_myIHSsentence_SINGLE_1}  Functional Form Sensitivity Analysis -- State Courts in Large Urban Counties}
{\footnotesize
\begin{tabularx}{\linewidth}{lCCCCC}

\toprule
&{(1)}&{(2)}&{(3)}&{(4)}&{(5)} \tabularnewline \midrule
\multicolumn{1}{c}{ }& \multicolumn{2}{c}{{OLS}} & \multicolumn{1}{c}{{Ordered}} & \multicolumn{2}{c}{{Ordered}}  \tabularnewline \multicolumn{1}{c}{ }& \multicolumn{2}{c}{} & \multicolumn{1}{c}{{Probit}} & \multicolumn{2}{c}{{Heckman}}  \tabularnewline  \cmidrule(l{2pt}r{5pt}){2-3}  \cmidrule(l{2pt}r{5pt}){4-4}  \cmidrule(l{2pt}r{5pt}){5-6}   \addlinespace[-2ex] \tabularnewline
{}&{Jail}&{Prison}&{}&{Jail}&{Prison} \tabularnewline
\midrule \addlinespace[\belowrulesep]
\textbf{Sentencing Equation}&&&&& \tabularnewline
\addlinespace[1ex] Black&.145***&.0977***&&.145***&.0964*** \tabularnewline
&(.0195)&(.0295)&&(.0192)&(.03) \tabularnewline
\addlinespace[1ex] Hispanic&.202***&.148***&&.2***&.165*** \tabularnewline
&(.0356)&(.036)&&(.0387)&(.0382) \tabularnewline
\addlinespace[1ex] \textbf{Selection Equation}&&&&& \tabularnewline
\addlinespace[1ex] Black&&&-.00147&-.00147&-.00147 \tabularnewline
&&&(.0181)&(.018)&(.018) \tabularnewline
\addlinespace[1ex] Hispanic&&&.0419***&.0413***&.0413*** \tabularnewline
&&&(.0135)&(.0135)&(.0135) \tabularnewline
\addlinespace[1ex] \midrule \addlinespace[1ex] \(\overline{sentence}_{W}\)&3.28&53.7&&3.28&53.7 \tabularnewline
\addlinespace[1ex] \(\rho_3\)&&&&-.062&-.062 \tabularnewline
&&&&(.162)&(.162) \tabularnewline
\(\rho_4\)&&&&.379&.379 \tabularnewline
&&&&(.143)&(.143) \tabularnewline
p-value: \(\rho_3=\rho_4=0\)&&&&.052&.052 \tabularnewline
p-value: Exclusion Restrictions&&&.000&.000&.000 \tabularnewline
\addlinespace[1ex] p-value: \(\beta_{Black} = \beta_{Hispanic}\)&.144&.09&&.192&.025 \tabularnewline
p-value: \(\alpha_{Black} = \alpha_{Hispanic}\)&&&.021&.023&.023 \tabularnewline
\addlinespace[1ex] \(R^2\)&.22&.332&&& \tabularnewline
Observations&38,939&24,945&96,057&96,057&96,057 \tabularnewline
\bottomrule \addlinespace[\belowrulesep]

\end{tabularx}
\begin{flushleft}
\scriptsize \textbf{Notes}: *** denotes significance at 1\%, ** at 5\%, and * at 10\%. Standard errors are clustered at county level. The dependant variable in the sentencing equation is the inverse hyperbolic sine of sentence length in months. In the selection equation, the dependant variable is stage, which takes values 0, 1, 2, 3 or 4. All specifications include the following control variables: county dummies, year of arrest dummies, most serious arrest offence code dummies, second most serious arrest offence code dummies, dummies for categories of the count of arrest charges, age decile dummies, dummies for most serious prior arrest, prior failure to appear in court, most serious prior conviction and a dummy for prior adult felony conviction for a violent offense. The exclusion restrictions for the ordered probit selection model are leave-out county\(\times\)year means of the proportion of individuals who are last seen in stage 1 and stage 2. Given the 2-stage stratified sample design, SCPS-supplied weights are used in all analysis, thus yielding estimates for the 75 most populous counties in the month of May.
\end{flushleft}
}
\end{table}

\end{center}
\clearpage{ }

\subsection{\label{sec:MCsims1}Monte Carlo Simulations I -- The Selection-Corrected Sentencing Gap}
In this section I run a series of simulations\footnote{I used the simulation section of \citet{CL2007} as a basis for this section.} in order to highlight the joint importance of (i) differential racial progression through the CJS stages and (ii) the sign and size of sample selection in the CJS, in leading to a divergence in the estimated Black-white sentencing gap using OLS and an ordered Heckman approach.\\
The simulated sample size is 2,000 -- 1,000 white arrestees and 1,000 Black arrestees. I simulate $Severity$ -- a measure of offense severity -- and $Z$ -- the exclusion restriction for the ordered probit, to be independent standard normal variables. The errors $\xi$ and $\epsilon$ are created as standard bivariate normal variables with correlation $\rho$.\\
The DGP for the selection equation is given by:
\begin{equation}
s_i^{*} = \alpha_1 Black_i + \alpha_2 Severity_i + \alpha_3 Z_i + \xi_i
\end{equation}
In order to keep the same proportion of arrestees that progress to each stage constant across the different parameter specifications, I use percentiles of $s_i^{*}$ to determine the cutoffs that give rise to realizations of $s_i$ values. I choose the proportions 0.4, 0.1, 0.1 and 0.4 for stages 0,1,2 and 3 respectively, meaning that 40\% of arrestees will progress to the sentencing stage. This choice is made to align the simulation results with those of the actual data that I use in the body of the paper.
\begin{align}
  s_{i} &=     \begin{cases}
                  \, 0 \quad \text{if } -\infty    < s_{i}^{*} \leq s_{40}^{*}\\
                  \, 1 \quad \text{if } s_{40}^{*} < s_{i}^{*} \leq s_{50}^{*}\\
                  \, 2 \quad \text{if } s_{50}^{*} < s_{i}^{*} \leq s_{60}^{*}\\
                  \, 3 \quad \text{if } s_{60}^{*} < s_{i}^{*}  <   \infty
                \end{cases} \label{Eq:oheck_sel_sim}
\end{align}
Finally I specify the DGP for the sentencing equations as:
\begin{equation}
y_{i} =     \begin{cases}
                  \beta_0 + \beta_1 Black_i + \beta_2 Severity_i + \epsilon_{i} \qquad \, \, \, \text{if } s_{i}=3 \\
                  \text{missing} \qquad  \qquad  \qquad  \qquad  \qquad  \qquad \, \, \, \,  \quad \text{otherwise} .  \label{Eq:oheck_out_sim}
            \end{cases} 
\end{equation}
I set the baseline parameters as follows:\\
Selection ($s_i^{*}$): $\alpha_2 = \alpha_3 = 1$.  \\
Outcome ($y_i$): $\beta_0 = ln(40)$, $\beta_1 = .1$, $\beta_2 = 1$.\\
In the simulations below I only ever change two parameters. These parameters are $\rho$, which determines the extent to which sample selection is present, and $\alpha_1$, which governs the differential progression on Black arrestees through the CJS stages. A higher $\alpha_1$ value means that  a higher fraction of Black individuals will progress to the final, sentencing stage.

\begin{center}
  \begin{table}[htb] \centering
\newcolumntype{C}{>{\centering\arraybackslash}X}

\caption{\label{tab:global_simulations_2a}Monte Carlo Simulations -- Comparing OLS and Ordered Heckman Estimates}
{\footnotesize
\begin{tabularx}{\linewidth}{lCCCCCCC}

\toprule
&{(1)}&{(2)}&{(3)}&{(4)}&{(5)}&{(6)}&{(7)} \tabularnewline \midrule
\multicolumn{1}{c}{ }& \multicolumn{2}{c}{{\textbf{Parameter Choices}}} & \multicolumn{2}{c}{{\textbf{OLS}}} & \multicolumn{2}{c}{{\textbf{Ordered Heckman}}}  & \multicolumn{1}{c}{{\textbf{OLS Bias}}} \tabularnewline  \cmidrule(l{2pt}r{5pt}){2-3}  \cmidrule(l{2pt}r{5pt}){4-5} \cmidrule(l{2pt}r{5pt}){6-7} \cmidrule(l{2pt}r{5pt}){8-8} \tabularnewline
{}&{\(\bm{\rho}\)}&{\(\bm{\alpha_1}\)}&{\(\bm{\hat{\beta}_1}\)}&{\textbf{95\% Coverage Probability}}&{\(\bm{\hat{\beta}_1}\)}&{\textbf{95\% Coverage Probability}}&{\(\bm{\hat{\beta}_1^{OLS} - \hat{\beta}_1^{OH}}\)} \tabularnewline
\midrule \addlinespace[\belowrulesep]
\addlinespace[2ex] 1.)&.50&.50&.0261&.812&.0998&.950&-.0737 \tabularnewline
&&&(.0686)&&(.0693)&&(.0204) \tabularnewline
\addlinespace[.5ex] 2.)&.50&.20&.0709&.929&.1005&.948&-.0295 \tabularnewline
&&&(.0679)&&(.0678)&&(.0171) \tabularnewline
\addlinespace[.5ex] 3.)&.50&.10&.0856&.943&.1000&.946&-.0145 \tabularnewline
&&&(.0683)&&(.0682)&&(.0166) \tabularnewline
\addlinespace[.5ex] 4.)&.50&0&.1005&.951&.1002&.951&.0003 \tabularnewline
&&&(.0675)&&(.0672)&&(.0163) \tabularnewline
\addlinespace[2ex] 5.)&.25&.50&.0634&.918&.1003&.950&-.0370 \tabularnewline
&&&(.0709)&&(.0720)&&(.0156) \tabularnewline
\addlinespace[.5ex] 6.)&.25&.20&.0855&.943&.1003&.948&-.0148 \tabularnewline
&&&(.0700)&&(.0701)&&(.0100) \tabularnewline
\addlinespace[.5ex] 7.)&.25&.10&.0929&.952&.1003&.951&-.0075 \tabularnewline
&&&(.0694)&&(.0695)&&(.0089) \tabularnewline
\addlinespace[.5ex] 8.)&.25&0&.1010&.947&.1008&.947&.0001 \tabularnewline
&&&(.0705)&&(.0705)&&(.0084) \tabularnewline
\addlinespace[2ex] 9.)&.10&.50&.0856&.945&.1002&.950&-.0146 \tabularnewline
&&&(.0710)&&(.0722)&&(.0140) \tabularnewline
\addlinespace[.5ex] 10.)&.10&.20&.0944&.945&.1003&.945&-.0059 \tabularnewline
&&&(.0707)&&(.0709)&&(.0069) \tabularnewline
\addlinespace[.5ex] 11.)&.10&.10&.0973&.950&.1002&.951&-.0030 \tabularnewline
&&&(.0706)&&(.0708)&&(.0050) \tabularnewline
\addlinespace[.5ex] 12.)&.10&0&.0999&.950&.0998&.951&.0000 \tabularnewline
&&&(.0706)&&(.0707)&&(.0044) \tabularnewline
\addlinespace[2ex] 13.)&0&.50&.0998&.948&.0999&.949&-.0001 \tabularnewline
&&&(.0717)&&(.0730)&&(.0135) \tabularnewline
\addlinespace[.5ex] 14.)&0&.20&.1004&.950&.1004&.949&.0000 \tabularnewline
&&&(.0707)&&(.0710)&&(.0062) \tabularnewline
\addlinespace[.5ex] 15.)&0&.10&.1004&.950&.1004&.949&.0000 \tabularnewline
&&&(.0706)&&(.0707)&&(.0038) \tabularnewline
\addlinespace[.5ex] 16.)&0&0&.1009&.950&.1009&.950&-.0000 \tabularnewline
&&&(.0703)&&(.0704)&&(.0029) \tabularnewline
\bottomrule \addlinespace[\belowrulesep]

\end{tabularx}
\begin{flushleft}
\scriptsize \textbf{Notes}: Results based on 10,000 simulation runs. The target parameter -- \(\beta_1\) -- is .1 for all simulations. Columns 3 and 5 show the mean and bootrapped standard error of \(\hat{\beta}_1\) from the 10,000 simulation runs for an OLS and ordered Heckman regression respecitvely. Columns 4 and 6 show the coverage probabilities of the 95\% confidence intervals for the OLS and ordered Heckman regressions resepectively -- the proportion of simulation runs for which the 95\% confidence interval of \(\hat{\beta}_1\) included the true parameter value, \(\beta_1 = .1\). Column 7 shows the mean and bootstrapped standard error of the difference between the OLS and ordered Heckman estimates of \(\beta_1\).
\end{flushleft}
}
\end{table}

\end{center}

The simulation results confirm that it is necessary to have \textit{both} differential racial progression through the CJS stages ($\alpha_1 \neq 0$) and for the sentencing stage to represent a selected sample of arrestees ($\rho \neq 0 $), in order for the OLS and ordered Heckman estimates of the Black-white sentencing gap to diverge. In Case 1, where there is both positive selection and differential racial progression through the CJS stages, I show a large and significant negative bias of the OLS estimate of the Black-white sentencing gap.

When I specify there to be no differential racial progression, but I keep the sample selection parameter fixed ($\rho=.5$) -- as I do in Case 4 -- there is no difference between the OLS and ordered Heckman estimates.

Likewise, when I shut down sample selection (i.e. set $\rho = 0$), then even with differential racial progression, the OLS and ordered Heckman estimates coincide (Case 13 is the best example of this, Cases 14 and 15 are useful too.).

These intermediate cases make the final case somewhat moot. Here I shut down both differential racial progression \textit{and} sample selection. Nevertheless, I present it for completeness. Unsurprisingly, there is no difference between the two estimates for this final case.
\clearpage{ }

\subsection{\label{sec:MCsims2}Monte Carlo Simulations II -- Sample Selection and the Imputation Method}
The purpose of this section is to provide greater detail on the simulation exercise presented in Table \ref{tab:global_simulations_7a_slim} and discussed in Section \ref{sec:SSP}.
The simulated sample size is 2,000 -- 1,000 white arrestees and 1,000 Black arrestees. I simulate $Severity$ -- a measure of offense severity -- and $Z$ -- the exclusion restriction for the ordered probit, to be independent standard normal variables. The errors $\xi$ and $\epsilon$ are created as standard bivariate normal variables with correlation $\rho$.\\
The DGP for the selection equation is given by:
\begin{equation}
s_i^{*} = \alpha_1 Black_i + \alpha_2 Severity_i + \alpha_3 Z_i + \xi_i
\end{equation}
In order to keep the same proportion of arrestees that progress to each stage constant across the different parameter specifications, I use percentiles of $s_i^{*}$ to determine the cutoffs that give rise to realizations of $s_i$ values. I choose the proportions 0.20, 0.05, 0.05 and 0.70 for stages 0, 1, 2 and 3 respectively, meaning that 70\% of arrestees will progress to the sentencing stage. This choice is made to ensure that the imputed sentence length mass point of those who do not reach the sentencing stage is sufficiently far away from the median.
\begin{align}
  s_{i} &=     \begin{cases}
                  \, 0 \quad \text{if } -\infty    < s_{i}^{*} \leq s_{20}^{*}\\
                  \, 1 \quad \text{if } s_{20}^{*} < s_{i}^{*} \leq s_{25}^{*}\\
                  \, 2 \quad \text{if } s_{25}^{*} < s_{i}^{*} \leq s_{30}^{*}\\
                  \, 3 \quad \text{if } s_{30}^{*} < s_{i}^{*}  <   \infty
                \end{cases} \label{Eq:oheck_sel_sim}
\end{align}
Finally I specify the DGP for the sentencing equations as:
\begin{equation}
y_{i} =     \begin{cases}
                  \beta_0 + \beta_1 Black_i + \beta_2 Severity_i + \epsilon_{i} \qquad \, \, \, \text{if } s_{i}=3 \\
                  \text{missing} \qquad  \qquad  \qquad  \qquad  \qquad  \qquad \, \, \, \,  \quad \text{otherwise} .  \label{Eq:oheck_out_sim}
            \end{cases} 
\end{equation}
The two parameters I vary across simulation specifications are $\rho$ and $\alpha_1$ -- the parameters that respectively govern the sample selection problem and differential stadial progression by race. The other parameters remain fixed across all simulation specifications, are as follows:\\
Selection ($s_i^{*}$): $\alpha_2 = \alpha_3 = 1$. This means that there is differential stadial progression \\
Outcome ($y_i$): $\beta_0 = ln(40)$, $\beta_1 = .1$, $\beta_2 = 1$.\\

Table \ref{tab:global_simulations_7a} below is an extended version of Table \ref{tab:global_simulations_7a_slim}, presenting results for a wider range of quantile regression specifications.
\begin{center}
  \begin{table}[htb] \centering
\newcolumntype{C}{>{\centering\arraybackslash}X}

\caption{\label{tab:global_simulations_7a}Monte Carlo Simulations -- The Imputation Method for Sample Selection Problems}
{\footnotesize
\begin{tabularx}{\linewidth}{lCCCCCCCCCC}

\toprule
&{(1)}&{(2)}&{(3)}&{(4)}&{(5)}&{(6)}&{(7)}&{(8)}&{(9)}&{(10)} \tabularnewline \midrule
\multicolumn{1}{c}{ }& \multicolumn{2}{c}{{\textbf{Parameter}}} &  \multicolumn{1}{c}{ }& \multicolumn{6}{c}{{\textbf{Quantile Regression + Imputation Method}}} \tabularnewline \multicolumn{1}{c}{ }& \multicolumn{2}{c}{{\textbf{Choices}}} &  \multicolumn{6}{c}{ } \tabularnewline  \cmidrule(l{2pt}r{5pt}){2-3}  \cmidrule(l{2pt}r{5pt}){5-10}  \addlinespace[-2ex] \tabularnewline
{}&{\(\bm{\rho}\)}&{\(\bm{\alpha_1}\)}&{\(\bm{\hat{\beta}_1^{OLS}}\)}&{\(\bm{\hat{\beta}_1^{40}}\)}&{\(\bm{\hat{\beta}_1^{50}}\)}&{\(\bm{\hat{\beta}_1^{60}}\)}&{\(\bm{\hat{\beta}_1^{70}}\)}&{\(\bm{\hat{\beta}_1^{80}}\)}&{\(\bm{\hat{\beta}_1^{90}}\)}&{\(\bm{\hat{\beta}_1^{OH}}\)} \tabularnewline
\midrule \addlinespace[\belowrulesep]
\addlinespace[2ex] 1.)&1&.50&-.002&.333&.267&.212&.170&.141&.120& \tabularnewline
&&&(.048)&(.097)&(.083)&(.074)&(.070)&(.071)&(.080)& \tabularnewline
\addlinespace[.5ex] 2.)&1&.20&.059&.178&.160&.141&.126&.116&.108& \tabularnewline
&&&(.048)&(.091)&(.080)&(.073)&(.069)&(.070)&(.080)& \tabularnewline
\addlinespace[.5ex] 3.)&1&.10&.079&.128&.124&.118&.111&.108&.104& \tabularnewline
&&&(.048)&(.089)&(.080)&(.072)&(.069)&(.070)&(.080)& \tabularnewline
\addlinespace[.5ex] 4.)&1&0&.100&.077&.089&.095&.097&.100&.101& \tabularnewline
&&&(.048)&(.089)&(.079)&(.073)&(.069)&(.070)&(.080)& \tabularnewline
\addlinespace[2ex] 5.)&.50&.50&.049&.450&.365&.293&.238&.196&.161&.100 \tabularnewline
&&&(.053)&(.117)&(.097)&(.083)&(.077)&(.077)&(.085)&(.054) \tabularnewline
\addlinespace[.5ex] 6.)&.50&.20&.079&.227&.200&.174&.154&.138&.124&.099 \tabularnewline
&&&(.053)&(.107)&(.093)&(.082)&(.077)&(.077)&(.085)&(.053) \tabularnewline
\addlinespace[.5ex] 7.)&.50&.10&.090&.154&.146&.136&.126&.119&.112&.100 \tabularnewline
&&&(.052)&(.106)&(.092)&(.081)&(.076)&(.076)&(.085)&(.052) \tabularnewline
\addlinespace[.5ex] 8.)&.50&0&.100&.079&.091&.096&.097&.100&.100&.100 \tabularnewline
&&&(.052)&(.104)&(.091)&(.081)&(.076)&(.076)&(.084)&(.053) \tabularnewline
\addlinespace[2ex] 9.)&.25&.50&.074&.489&.398&.324&.267&.224&.186&.100 \tabularnewline
&&&(.053)&(.124)&(.101)&(.086)&(.078)&(.077)&(.086)&(.054) \tabularnewline
\addlinespace[.5ex] 10.)&.25&.20&.089&.244&.214&.187&.166&.149&.134&.099 \tabularnewline
&&&(.053)&(.112)&(.095)&(.083)&(.078)&(.078)&(.087)&(.054) \tabularnewline
\addlinespace[.5ex] 11.)&.25&.10&.095&.162&.153&.142&.132&.124&.117&.100 \tabularnewline
&&&(.054)&(.110)&(.094)&(.084)&(.078)&(.078)&(.086)&(.054) \tabularnewline
\addlinespace[.5ex] 12.)&.25&0&.100&.080&.091&.096&.099&.100&.100&.100 \tabularnewline
&&&(.053)&(.110)&(.094)&(.084)&(.078)&(.077)&(.086)&(.053) \tabularnewline
\addlinespace[2ex] 13.)&0&.50&.100&.519&.426&.350&.294&.250&.212&.100 \tabularnewline
&&&(.054)&(.126)&(.102)&(.087)&(.080)&(.079)&(.087)&(.055) \tabularnewline
\addlinespace[.5ex] 14.)&0&.20&.100&.255&.225&.197&.177&.160&.146&.100 \tabularnewline
&&&(.054)&(.115)&(.096)&(.084)&(.078)&(.077)&(.087)&(.054) \tabularnewline
\addlinespace[.5ex] 15.)&0&.10&.100&.169&.159&.146&.138&.130&.122&.100 \tabularnewline
&&&(.054)&(.112)&(.095)&(.085)&(.079)&(.078)&(.086)&(.054) \tabularnewline
\addlinespace[.5ex] 16.)&0&0&.100&.081&.092&.095&.099&.100&.100&.100 \tabularnewline
&&&(.054)&(.113)&(.096)&(.085)&(.079)&(.078)&(.087)&(.054) \tabularnewline
\bottomrule \addlinespace[\belowrulesep]

\end{tabularx}
\begin{flushleft}
\scriptsize \textbf{Notes}: Results based on 10,000 simulation runs. The target parameter -- \(\beta_1\) -- is .1 for all simulations. The other fixed parameters are: \(\alpha_2=\alpha_3= \beta_2=1\) and  \( \beta_0=ln(40)\). Column 3 shows the mean and bootrapped standard error of \(\hat{\beta}_1\) from an OLS regression based on the sentencing stage alone. Columns 4-9 present the mean and bootrapped standard error of \(\hat{\beta}_1\) from quantile regressions for the 40th-90th percentiles in decile increments based on an imputation approach whereby all missing sentences are allocated the lowest value of sentence length in each iteration run. Finally, column 10 shows the mean and bootrapped standard error of \(\hat{\beta}_1\) from an ordered Heckman regression.
\end{flushleft}
}
\end{table}

\end{center}
\,
\clearpage{ }
\,
\beginappendixB
\section{\label{sec:AppendixData}Data Appendix}
In both the federal and state datasets, I link across multiple stages of the criminal justice system. In this Appendix section I briefly outline the respective processes.

\subsection{Federal Justice Statistics Program (FJSP) Data Linkage}

The report by \cite{Kelly2012} was invaluable in conducting the merges across the four federal agency files used for the analysis. These agency files are part of the Federal Justice Statistics Program (FJSP), and are made available as individual Standard Analysis Files (SAF) -- standardized files at the individual-case level.

The four sets of agency datasets/SAFs comprise i.) data from the U.S. Marshals Service (USMS) that covers arrests, ii.) data from the Executive Office for United States Attorneys (EOUSA) that covers, amongst other things, case filing, iii.) data from the Administrative Office of the United States Courts (AOUSC) covering charging and iv.) data from the United States Sentencing Commission (USSC) that covers sentencing.

It is possible to link files inter-agency (e.g., USMS In to EOUSA Matters Out), as well as intra-agency (EOUSA Matters Out to EOUSA Cases Out). All linkage files are dyadic, which means in order to link the USMS arrest data to USSC sentencing outcomes, I need to go via the EOUSA and AOUSC agency files. 

I refer the reader interested in learning more about the data linkages, and the lengths taken to validate the linking process, to the work of \cite{Kelly2012}.

In constructing my main stage variable ($s_i$ in Equation \ref{Eq:oheck_sel}), I use information from each of the four agency files, as well as from the linking files. If an individual is seen only in the USMS data (USMS IN), but either a.) not in the EOUSA data (EOUSA Matters Out) or b.) has been removed by the data providers from the EOUSA based on a screening algorithm (in this case if the individual has an arrest code that relates to material witnesses and supervision violations) then  I allocate the individual a value of $s_i = 0$. For those seen in the USMS and EOUSA data, but are not present in the AOUSC data (AOUSC Cases Out), the ascribe $s_i = 1$. For those last seen in the AOUSC data (i.e., are not linked to the USSC data(USSC Out)) I code $s_i = 2$. For those linked across all stages, I code $s_i = 3$.\footnote{The data provider applies another screening algorithm at this linkage, removing any individuals who were not convicted of a charge who appear in the USSC data.}\footnote{For the extension in Section \ref{sec:ResultsFCJSext}, I split those with $s_i = 3$ into $s_i = 3$ for those who are sentenced without a statutory minimum charge, and $s_i = 4$ for those who face a statutory minimum sentence charge.}

\subsection{New Orleans District Attorney's Office (NODA) Data Linkage}
The data linkage for the New Orleans state data is considerably more straightforward. There are multiple datasets, including separate files for arrest outcomes, charging details, assistant DA characteristics, judge characteristics, defendant characteristics, and sentencing outcomes. The data are at different levels e.g. charging data is at the defendant-offense-charge level, whereas the assistant DA data is at the attorney level. A series of unique identifiers enables the linkage across both datasets and levels.

\subsection{State Court Processing Statistics (SCPS) Data Linkage}
The SCPS data that I use for analysis in this paper was obtained from the ICPSR\footnote{See this link -- \url{https://www.icpsr.umich.edu/web/ICPSR/studies/2038} -- for more information on the data.} pre-linked.

\end{document}